\documentclass[longauth]{aa}
\usepackage{subfigure}
\usepackage{graphicx}
\usepackage{txfonts}
\usepackage{upgreek}
\usepackage{natbib}
\usepackage{xcolor}
\usepackage[colorlinks=true,linkcolor=blue,citecolor=blue,urlcolor=blue]{hyperref}

\def\as {\ifmmode {\rlap.}$\,$''$\,$\! \else ${\rlap.}$\,$''$\,$\!$\fi}

\begin{document} 

\title{ANTIHEROES-PRODIGE: Quantifying the connection from envelope to disk with the IRAM 30m telescope and NOEMA}
\subtitle{I. Attack of the streamers: L1448N's fight for order in the chaos}
\titlerunning{ANTIHEROES-PRODIGE: Quantifying the connection from envelope to disk}
\author{
	C. Gieser\inst{1}
	\and
	P. Caselli\inst{2}
	\and
	D.~M. Segura-Cox\inst{3, 2}	
	\and
	J.~E. Pineda\inst{2}
	\and 
	L.~A. Busch\inst{2}
	\and
	M.~T. Valdivia-Mena\inst{4}
	\and
	M.~J. Maureira\inst{2}
	\and
	Y. Lin\inst{2}
	\and
	T.~H. Hsieh\inst{5,6}
	\and
	Y.~R. Chou\inst{2}
	\and
	L. Bouscasse\inst{7}
	\and
	P.~C. Cortés\inst{8,9}
	\and
	N. Cunningham\inst{10}
	\and
	A. Dutrey\inst{11}
	\and
	A. Fuente\inst{12}
	\and
	Th. Henning\inst{1}
	\and
	A. Lopez-Sepulcre\inst{7,13}
	\and
	J.~J. Miranzo-Pastor\inst{12}
	\and
	R. Neri\inst{7}
	\and
	D. Semenov\inst{14,1}
	\and
	M. Tafalla\inst{15}
	\and
	S.~E. van Terwisga\inst{16}
	}

	\institute{
	Max-Planck-Institut für Astronomie, Königstuhl 17, D-69117 Heidelberg, Germany\\ \email{gieser@mpia.de}
	\and
	Max-Planck-Institut für extraterrestrische Physik, Giessenbachstrasse 1, 85748, Garching, Germany
	\and
	Department of Physics and Astronomy, University of Rochester, Rochester, NY, 14627, USA 
	\and
	European Southern Observatory, Karl-Schwarzschild-Strasse 2 85748 Garching bei München, München, Germany
	\and
	Taiwan Astronomical Research Alliance (TARA), Taiwan
	\and
	Institute of Astronomy and Astrophysics, Academia Sinica, No. 1, Sec. 4, Roosevelt Road, Taipei 10617, Taiwan
	\and
	Institut de Radioastronomie Millimétrique (IRAM), 300 rue de la Piscine, F-38406, Saint-Martin d’Hères, France
	\and
	Joint ALMA Observatory, Alonso de Córdova 3107, Vitacura, Santiago, Chile
	\and
	National Radio Astronomy Observatory, 520 Edgemont Road, Charlottesville, VA 22903, USA
	\and
	SKA Observatory, Jodrell Bank, Lower Withington, Macclesfield SK11 9FT, UK
	\and
	Laboratoire d’astrophysique de Bordeaux, Univ. Bordeaux, CNRS, B18N, allée Geoffroy Saint-Hilaire, 33615 Pessac, France
	\and
	Centro de Astrobiología (CAB), CSIC-INTA, Ctra. de Torrejón a
Ajalvir, km 4, 28850 Torrejón de Ardoz, Spain
	\and
	IPAG, Universit\'{e} Grenoble Alpes, CNRS, F-38000 Grenoble, France
	\and
	Zentrum für Astronomie der Universität Heidelberg, Institut für Theoretische Astrophysik, Albert-Ueberle-Str. 2, 69120 Heidelberg, Germany
	\and
	Observatorio Astronómico Nacional (IGN), Alfonso XII 3, 28014 Madrid, Spain
	\and
	Space Research Institute, Austrian Academy of Sciences, Schmiedlstr. 6, 8042 Graz, Austria
	}

	\date{Received x; accepted x}

	\abstract
	{Star formation is a hierarchical process ranging from molecular clouds down to individual protostars. In particular how infalling asymmetric structures, called streamers, delivering new material onto protostellar systems, are connected to the surrounding envelope is not understood.}
	{We aim to investigate the connection between the cloud material at 10\,000\,au scales down to 300\,au scales towards L1448N in the Perseus star-forming region hosting three young Class 0/I protostellar systems (IRS3A, IRS3B, and IRS3C).}
	{Sensitive molecular line observations taken with the Institut de Radioastronomie Millimetrique (IRAM) 30m telescope and the Northern Extended Millimeter Array (NOEMA) at 1.4\,mm are used to study the kinematic properties in the region traced by the molecular lines (C$^{18}$O, SO, DCN, and SO$_{2}$). The temperature in the region is estimated using transitions of c-C$_{3}$H$_{2}$.}
	{Several infalling streamers are associated with the protostellar systems, some of them traced by C$^{18}$O and DCN, while one of them is bright in SO and SO$_2$. The kinematic properties of the former streamers are consistent with the velocities observed at large envelope scales of 10\,000 au, while the latter case show different kinematics. The masses and infall rates of the streamers are 0.01\,$M_\odot$ and 0.01-0.1\,$M_\odot$ and 10$^{-6}$\,$M_\odot$\,yr$^{-1}$ and $5-18\times10^{-6}$\,$M_\odot$\,yr$^{-1}$ for IRS3A and IRS3C, respectively. The envelope mass in the L1448N region is $\approx$16\,$M_\odot$, thus the mass of a single streamer is low compared to the envelope mass ($<$1\%). However, compared to the estimated mass of the protostellar systems a single streamer could deliver up to 1\% and 8-17\% of mass towards IRS3A ($M_* \approx 1.2$\,$M_\odot$) and IRS3C ($M_* \approx 1$\,$M_\odot$), respectively.}
	{The rotational signatures of structures in L1448N are all connected - from the large-scale envelope, infalling streamers, down to the rotation of all three disks. Two of the three Class 0/I protostellar systems are still fed by this surrounding material, which can be associated to the remnant envelope. However, we also find a streamer that is bright in SO and SO$_{2}$ towards IRS3C that could be connected to a nearby sulfur reservoir. Further studies are required to study the diverse chemical compositions and the origin of the streamers.}
	


	\maketitle

\section{Introduction}\label{sec:intro}

	Investigating the earliest and most embedded stages of star formation is essential for understanding (proto)stellar evolution, the properties of surrounding protoplanetary disks, and the formation of embedded planets. In the Class 0/I stage during low-mass star formation, a protostellar system is still surrounded by a large mass reservoir that allows further mass growth \citep{Evans2009, Dunham2014}. Understanding how the envelope material feeds protostellar systems requires multi-scale observations from large envelope to small disk scales \citep[][]{Pineda2023}.
	
\begin{figure*}[!htb]
\centering
\includegraphics[width=0.8\textwidth]{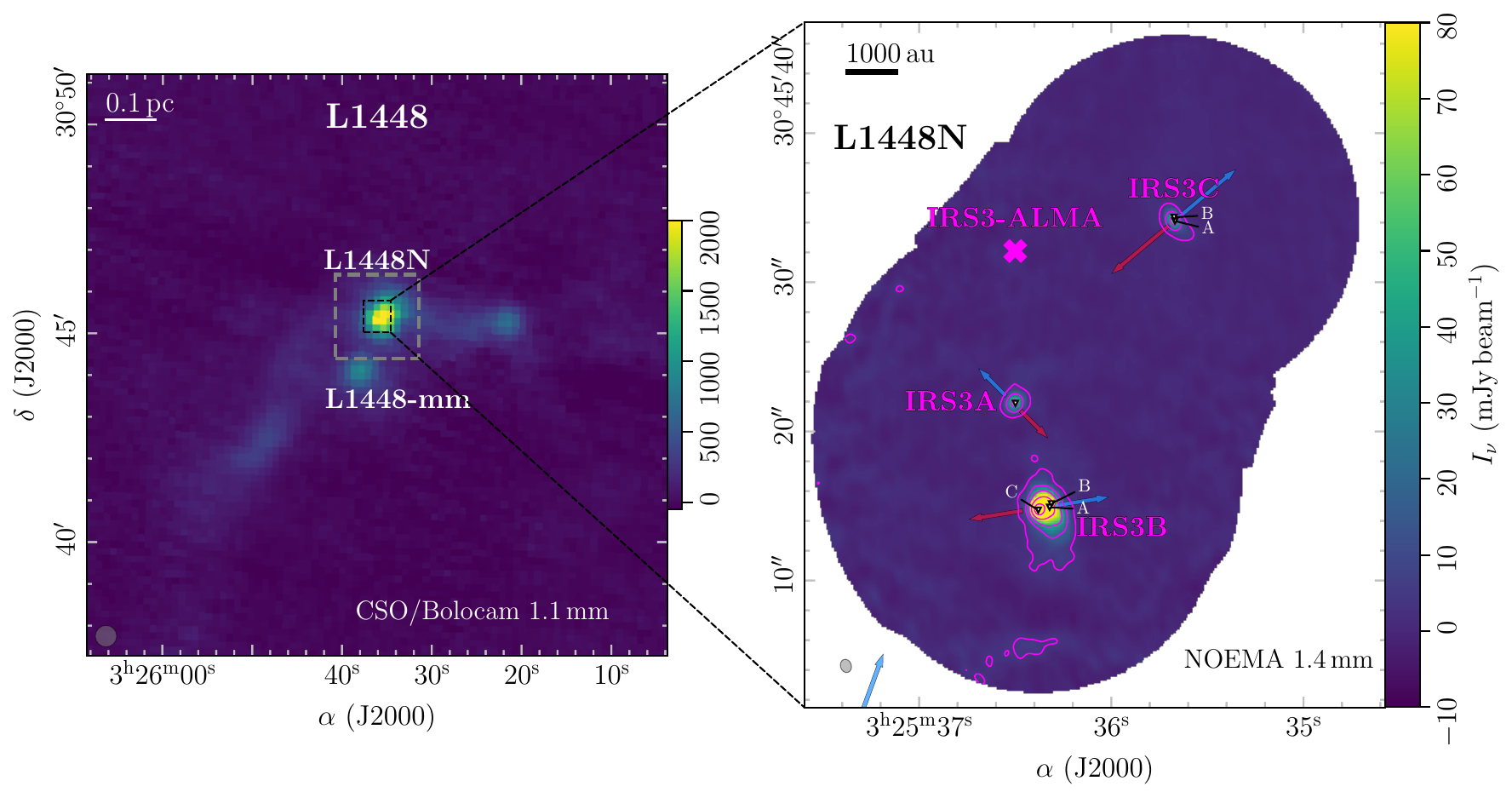}
\caption{Continuum images toward L1448N. The left panel shows large-scale 1.1\,mm continuum emission of L1448 \citep{Enoch2006}. The beam size is shown in the bottom left. The map size of the IRAM\,30m observations is highlighted by the grey dashed rectangle. The right panel shows the NOEMA 1.4\,mm continuum of L1448N with contour levels at 5, 25, 60, 120, 200$\times \sigma_\mathrm{cont}$ with $\sigma_\mathrm{cont}$=0.78\,mJy\,beam$^{-1}$. The triangles and white labels mark the positions of individual protostars \citep{Tobin2016} and the pink cross marks the recently detected IRS3-ALMA source \citep{Looney2025}. Bipolar outflow directions launched by the three protostellar systems are marked by red and blue arrows, including the blueshifted outflow lobe launched by L1448-mm towards the southeast. The synthesized beam of the NOEMA continuum data is shown in the bottom left corner. In both panels scale bars are shown in the top left corner.}
\label{fig:continuum}
\end{figure*}	

	A spherically symmetric collapse of an isolated core was first considered in early theoretical models of the formation of a Solar-type star \citep[e.g.,][]{Larson1969, Penston1969, Shu1977}. Sensitive and high spatial resolution observations from infrared to radio wavelengths over the past decades have revealed that star formation is a complex process, where the environment and asymmetries have to be taken into account, for example, filamentary structures in molecular clouds \citep{Andre2014, Hacar2023}, multiplicity \citep{Offner2023}, protoplanetary disks \citep{Miotello2023}, chemical diversity \citep{Ceccarelli2023}, and streamers \citep[][]{Pineda2020}. In this work, we use the definition that a streamer is an asymmetric structure with kinematically confirmed infall motions. More recent models and simulations including, for example, turbulence and magnetic fields, also show asymmetries at all scales relevant for star formation \citep[e.g.,][]{Walch2009, Offner2010, BallesterosParedes2015, Seifried2015, Chira2018, Kuffmeier2019, Kuznetsova2022, Kuffmeier2024}.
	
	Asymmetric infalling streamers have been identified in interferometric data in various molecular lines, for example, HC$_{3}$N \citep{Pineda2020,ValdiviaMena2024}, CO \citep{Ginski2021}, $^{13}$CO \citep{Gupta2024}, C$^{18}$O \citep{Yen2014,Flores2023, Kido2023, Mercimek2023, Cacciapuoti2024, Kido2025}, HCO$^{+}$ \citep{Yen2019,Garufi2022}, C$_{2}$H \citep{Tanious2024}, H$_{2}$CO \citep{ValdiviaMena2022,Podio2024}, CH$_{3}$CN \citep{FernandezLopez2023} as well as deuterated species such as DCN \citep[][Cortes et al. subm.]{Hsieh2023, Gieser2024}. The origin and effects of these infalling structures, including mass infall rates and potential shocks at the landing sites in the disk, have yet to be thoroughly investigated. Most importantly, the connection between streamers and the surrounding envelope material remains unclear, as the investigation is challenged by the limited field of view (FOV) and insufficient line sensitivity of current interferometers.

	Recently, a double infalling streamer structure, referred to as the bridge, has been found towards the IRS3A protostar located in L1448N \citep{Gieser2024}. The L1448N star-forming complex is located in the Perseus molecular cloud at a distance of $288\pm6\pm13$\,pc \citep[with statistical and systemic uncertainties, respectively,][]{Zucker2018}. This region contains three Class 0/I protostellar systems (IRS3A, IRS3B, and IRS3C) which host a single, triple, and binary system, respectively \citep{Tobin2018}. While IRS3B and IRS3C are Class 0 systems, in the case of IRS3A it is not clear if it is in the Class 0 or I stage \citep[e.g.,][]{Ciardi2003}, hence in the following we refer to all systems as Class 0/I. High angular resolution continuum observations with the Atacama Large Millimeter/submillimeter Array (ALMA) have revealed a ring and spiral-arm structures in the IRS3A and IRS3B disks, respectively, while the two protostars in the IRS3C binary system have each a compact disk \citep{Tobin2016, Reynolds2021, Reynolds2024}. An overview of the dust continuum distribution of L1448N is presented in Fig. \ref{fig:continuum} including the orientation of the outflows in the systems \citep{Lee2016,Reynolds2021,Gieser2024,Dunham2024}. The outflow in the IRS3B multiple system can be associated with the IRS3B-C protostar \citep{Reynolds2021}, while for IRS3C, currently available data do not allow us to distinguish which protostar in the binary drives the outflow \citep{Tobin2018}.
	
	The three Class 0/I systems of L1448N are target regions of the PROtostars \& DIsks: Global Evolution (PRODIGE) survey, a large program with the Northern Extended Millimeter Array (NOEMA) as part of the Max-Planck IRAM Observatory program (MIOP). The NOEMA continuum and molecular line observations at 1.4\,mm probe spatial scales down to 300\,au at the distance of the Perseus molecular cloud (1$''$ angular resolution), tracing the environment of the protostellar systems. PRODIGE observations of DCN ($3-2$) revealed an extended gas bridge surrounding the IRS3A and IRS3B systems that is infalling onto IRS3A \citep{Gieser2024}. The gas bridge showed an even larger extent in other molecular line tracers such as C$^{18}$O ($2-1$) and H$_{2}$CO ($3_{0,3}-2_{0,2}$). However, spatial filtering in the NOEMA observations due to missing short baseline data prevented studying the connection of the gas bridge to the environment as well as estimating masses and infall rates of the streamers.
	
	To complement the NOEMA data of PRODIGE, we are carrying out the IRAM 30\,m telescope large program
	``Qu\textbf{anti}fying t\textbf{h}e conn\textbf{e}ction f\textbf{r}om envel\textbf{o}p\textbf{e} to di\textbf{s}k with MIOP-PRODIGE'' (ANTIHEROES). The sensitive single-dish maps allow us to study the large scale properties of the molecular gas surrounding the protostellar systems from up to a few 10\,000\,au down to 3\,600\,au (12$''$). The combination of the interferometric and single-dish observations allow us to recover the flux on all scales, and thus we are able to reliably estimate molecular column densities and kinematic properties down to 300\,au scales for the entire PRODIGE sample. 
	
	In this work, we analyze the connection of infalling streamers to the larger scale material, from 36\,000\,au (120$''$) down to 300\,au (1$''$) scales, using a combination of IRAM\,30m telescope and NOEMA molecular line observations to demonstrate the potential of the combined data. The paper is organized as follows: Section \ref{sec:observations} contains a description of the PRODIGE and ANTIHEROES data reduction, as well as the combination of both data sets. The analysis of the molecular line data is presented in Sect. \ref{sec:results} and our results are further discussed in Sect. \ref{sec:discussion}. Our conclusions are summarized in Sect. \ref{sec:conclusions}.

\section{Observations}\label{sec:observations}

	In this section, we explain the data reduction of both the NOEMA and IRAM\,30m data sets and the merging of the data. The molecular line properties (quantum numbers, rest-frequency $\nu$, upper energy level $E_\mathrm{u}/k_\mathrm{B}$, Einstein A-coefficient $A_\mathrm{ul}$) of all transitions analyzed in this work, as well as the angular resolution and noise properties of the observations, are summarized in Table \ref{tab:obs}.

\begin{figure*}[!htb]
\centering
\includegraphics[width=0.7\textwidth]{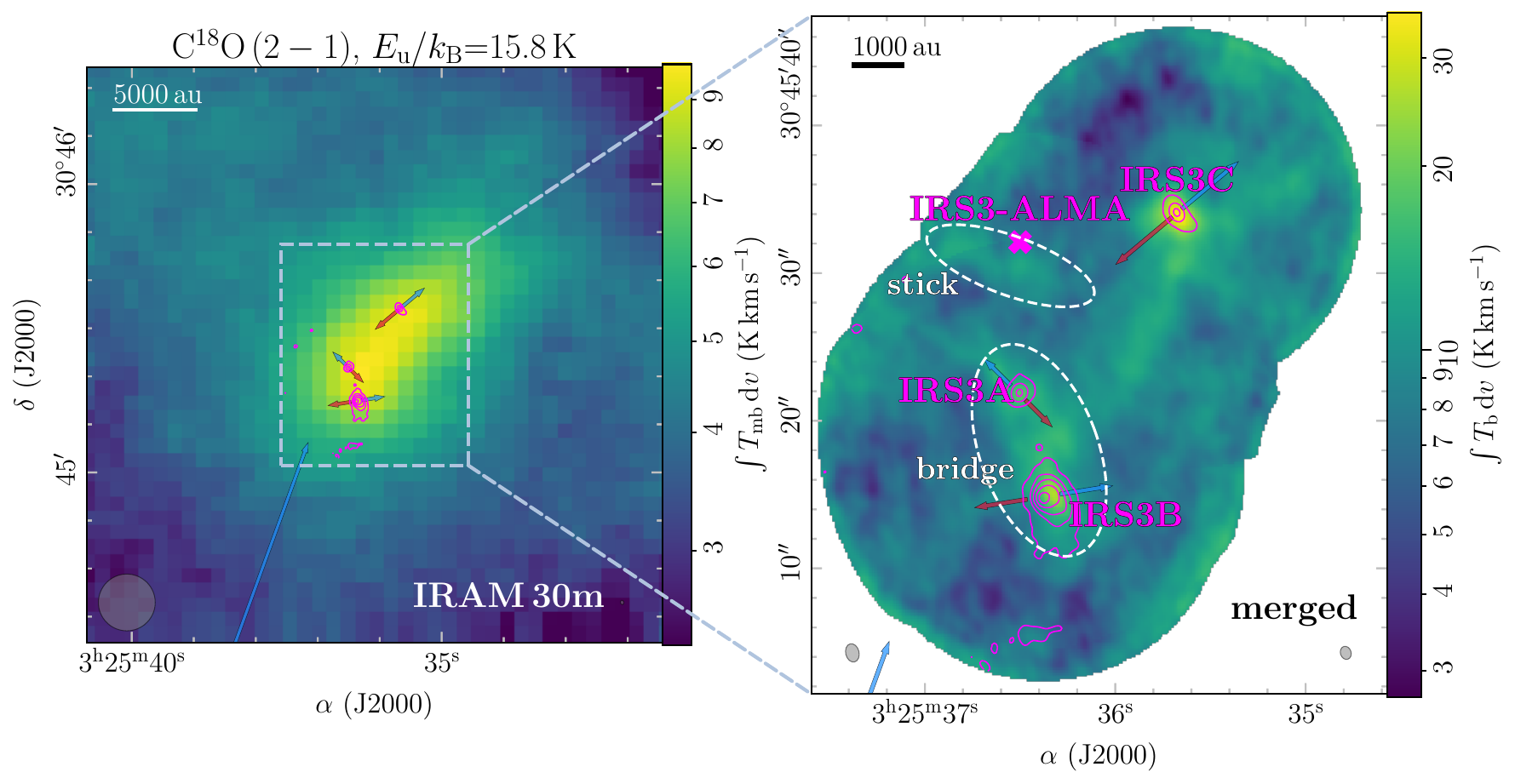}
\includegraphics[width=0.7\textwidth]{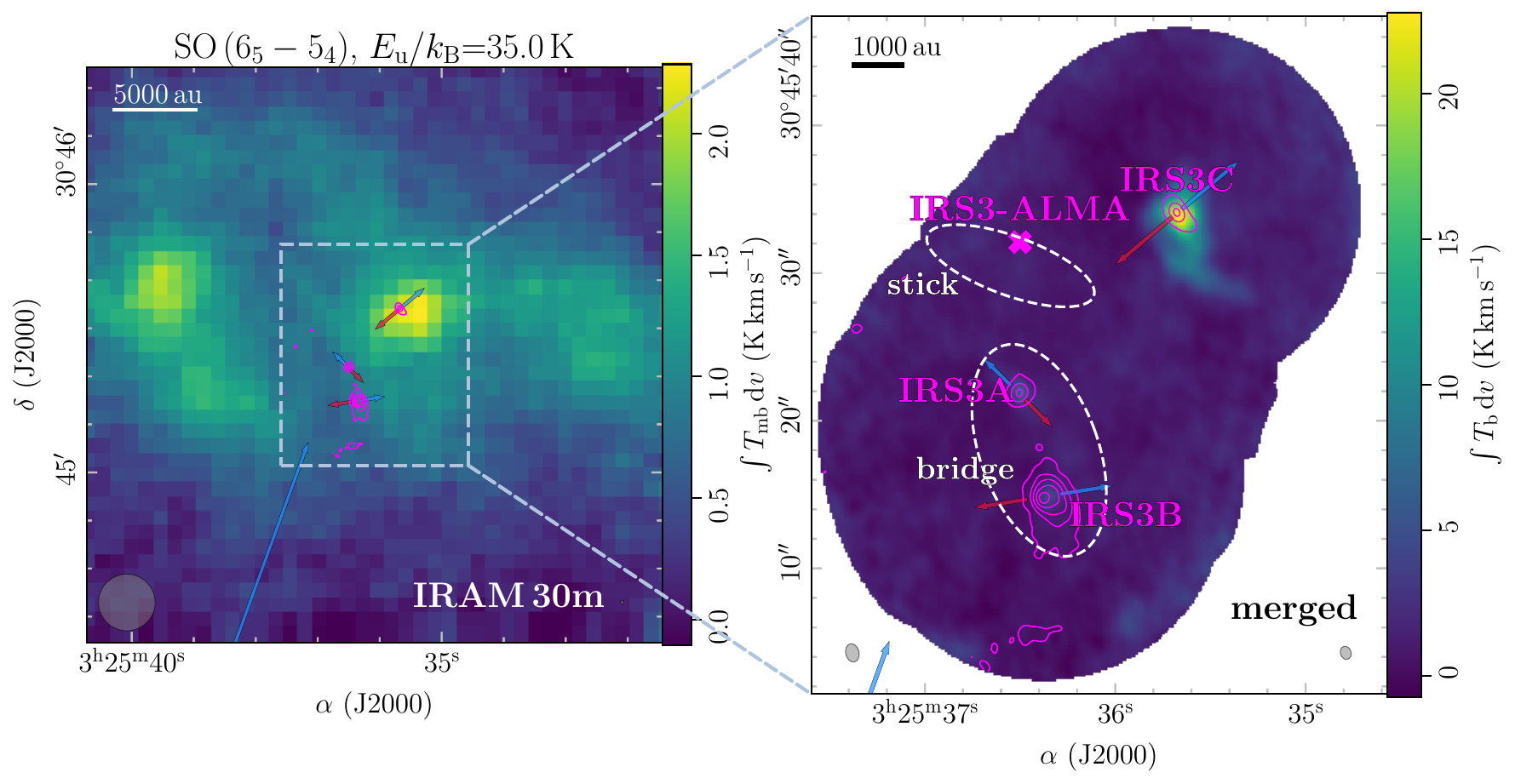}
\includegraphics[width=0.7\textwidth]{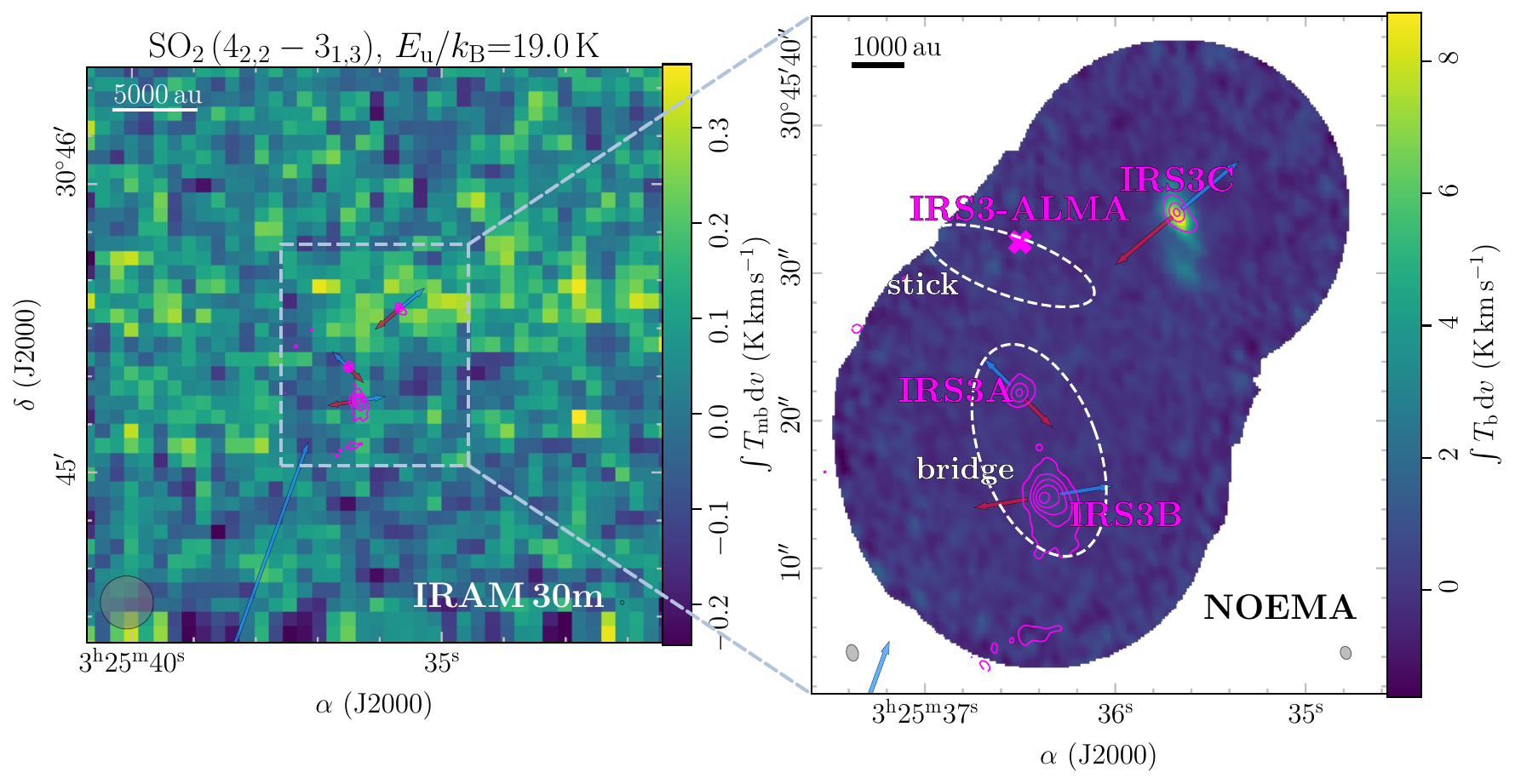}
\caption{Line integrated intensity map of C$^{18}$O ($2-1$), SO ($6_{5}-5_{4}$), and SO$_{2}$ ($4_{2,2}-3_{1,3}$) of the IRAM\,30m data (left) and merged or NOEMA data (right). In all panels the line integrated intensity is shown in color. Contours and arrows are the same as in Fig. \ref{fig:continuum}. The synthesized beam of the line and continuum data is shown in the bottom left and right corner, respectively. In all panels scale bars are shown in the top left corner. In the right panels, distinct structures (bridge and stick) seen in molecular emission are indicated by dashed ellipses.}
\label{fig:C18OSOSO2mom0}
\end{figure*}	

	\subsection{NOEMA continuum and molecular line data}
	
	The PRODIGE project (program id: L19MB, PIs: P. Caselli and Th. Henning) covers L1448N with NOEMA in the 1.4\,mm band. Three individual pointings toward the IRS3A, IRS3B, and IRS3C protostellar systems were obtained from December 2019 to October 2022 in the C- and D-array configurations. The projected baselines range from 15.9\,m up to 402\,m. The frequency coverage of the wideband data is $214.7-222.8$\,GHz in the lower sideband (LSB) and $230.2-238.3$\,GHz in the upper sideband (USB) at a channel width of 2\,MHz ($\approx$2.7\,km\,s$^{-1}$ at 1.4\,mm). A total of 39 high spectral resolution units with smaller bandwidths (narrowband data) were placed within the LSB and USB frequency ranges with a channel width of 62.5\,kHz ($\approx$0.09\,km\,s$^{-1}$ at 1.4\,mm). A summary of all molecular transitions covered by the narrowband data that were analyzed in this work is presented in Table \ref{tab:obs}.
	
	 A detailed description of the NOEMA data calibration procedure using \texttt{CLIC}\footnote{\texttt{CLIC}, \texttt{CLASS}, and \texttt{MAPPING} are software from the standard observatory pipeline in the Grenoble Image and Line Data Analysis Software (\texttt{GILDAS}, \url{https://www.iram.fr/IRAMFR/GILDAS/})}, including phase self-calibration using \texttt{MAPPING}, of the IRS3A and IRS3B PRODIGE data is presented in Sect. 2.1 in \citet{Gieser2024}. For the purpose of this work, we applied the same methodology to the additional IRS3C pointing. The three NOEMA pointings have sufficient overlap within their primary beam in order to combine all pointings into a mosaic \citep[as described in Sect. 2.2 in][]{Gieser2024}. 
	 
	Using \texttt{MAPPING}, the continuum and SO$_{2}$ NOEMA data were imaged with robust weighting (robust parameter of 1) and natural weighting, respectively, and cleaned with the Hogbom algorithm \citep{Hogbom1974} given that the emission is compact (Figs. \ref{fig:continuum} and \ref{fig:C18OSOSO2mom0}). Given the non-detection of SO$_{2}$ in the IRAM\,30m data (Fig. \ref{fig:C18OSOSO2mom0}), we only use the NOEMA data for this molecular transition, as only noise would be added in the data merging. For all other transitions analyzed in this work (Table \ref{tab:obs}), extended emission is detected in the IRAM\,30m data. Hence for these transitions, merging both data sets is crucial for the recovery of missing flux from spatial filtering. The remaining molecular line data were first merged with the IRAM\,30m data, imaged with natural weighting, and then cleaned with the SDI algorithm \citep{Steer1984} that takes into account the extended emission (as described further in Sect. \ref{sec:merging}). To all data, primary beam correction was applied with a primary beam size of $\approx22''$ at 1.4\,mm. To increase the signal-to-noise ratio (S/N) of the spectral line data, we rebinned the narrowband data to a common channel width of 0.1\,km\,s$^{-1}$.
	 
	The continuum mosaic data were cleaned down to 3$\times$ the theoretical noise with a support mask around the emission of the three protostellar systems. The synthesized beam ($\theta_\mathrm{maj}\times\theta_\mathrm{min}$) of the continuum data is $0\as91\times0\as73$ with a position angle (PA) of 18$^\circ$. The continuum noise level, $\sigma_\mathrm{cont}$, is 0.78\,mJy\,beam$^{-1}$ estimated from an emission-free area. The SO$_{2}$ NOEMA data were cleaned down to the theoretical noise and no support mask. The imaging and cleaning of the remaining molecular line data that were first merged with the IRAM 30m data are further presented in Sect. \ref{sec:merging}.
	
	\subsection{IRAM\,30m telescope line data}
	
	To avoid missing extended line emission that may be resolved out by the interferometer, we combined the NOEMA data with complementary zero-spacing spectral line observations from the IRAM 30m telescope using the Eight MIxer Receiver (EMIR) covering the same frequency range as the NOEMA data (program code 096-23, PI: C. Gieser). These observations of L1448N are the pilot project for the ongoing ANTIHEROES large program (program code 162-24, PI: C. Gieser) that includes the remaining protostellar systems of the entire PRODIGE sample. The fast Fourier transform spectrometer (FTS) backend provides a channel width of 50\,kHz ($\sim$0.07\,km\,s$^{-1}$).
	
	The IRAM\,30m observations of L1448N were carried out in the on-the-fly (OTF) mode in February and April 2024. The central position of the map is $\alpha_\mathrm{J2000}$=$3^h25^m36^s$ and $\delta_\mathrm{J2000}$=$30^\circ45'30''$. The OTF maps have a size of $120''\times120''$ and the region was observed in zigzag scanning mode (scanning speed 5$''$\,s$^{-1}$) in both right ascension and declination direction. The reference offset position ($\alpha_\mathrm{J2000}$=$3^h26^m40^s$ and $\delta_\mathrm{J2000}$=$30^\circ50'0''$) was selected based on the non-detection of $^{13}$CO emission in the COMPLETE survey data \citep{Ridge2006}. Pointing and focus settings were checked and corrected every $\approx$1\,h and $\approx$2\,h, respectively, on nearby and bright sources. 
	
	The data reduction of the single-dish data was carried out with the \texttt{CLASS} package. In individual spectra, baseline subtraction was carried out by fitting a first order polynomial around emission lines. The data were converted from antenna temperature ($T^*_\mathrm{A}$) to main beam temperature ($T_\mathrm{MB}$) using $T_\mathrm{MB} = \frac{F_\mathrm{eff}}{B_\mathrm{eff}}\times T^*_\mathrm{A}$ with forward efficiency $F_\mathrm{eff} = 0.94$ and main beam efficiency $B_\mathrm{eff} = 0.63$\footnote{\url{https://publicwiki.iram.es/Iram30mEfficiencies}}. The single-dish spectral line data were also rebinned to 0.1\,km\,s$^{-1}$ to match the spectral grid of the NOEMA data. Standalone IRAM\,30m data cubes were created using \texttt{xy\_map} with a pixel scale of 3$''$. The half power beam width (HPBW) is $\approx$12$''$ at 1.4\,mm corresponding to 3\,600\,au at the distance of L1448N.
	
	\subsection{Data merging of the line data}\label{sec:merging}
	
	We merged the interferometric and single-dish spectral line data using the \texttt{uv\_short} task in \texttt{MAPPING}. The merged line data were imaged with natural weighting and cleaned with no support mask down to the theoretical noise using the SDI algorithm. As a second threshold we set a maximum number of 5000 clean components per channel to avoid overcleaning channels with very extended emission. The pixel scale was set to 0.15$''$ and the angular resolution is $\approx$1$''$ (300\,au at the distance of L1448N).
				
	The channel width $\delta \varv$, beam size and noise level of the IRAM\,30m ($\theta$, $\sigma_\mathrm{line,30m}$) and merged or NOEMA ($\theta_\mathrm{maj}\times\theta_\mathrm{min}$ (PA), $\sigma_\mathrm{line}$) spectral line data are summarized for all transitions in Table \ref{tab:obs}. The line sensitivity is $\approx$100\,mK and $\approx$0.4\,K for the IRAM\,30m and merged data, respectively. This data set provides sensitive molecular line observations from scales of 36\,000\,au (120$''$) down to 300\,au (1$''$) which is crucial to understand the origin of molecular gas, such as streamers, that feed the protostellar systems. The importance of merging the IRAM\,30m data with the NOEMA observations is further evaluated in Appendix \ref{sec:app_merging} comparing the spatial distribution as well as individual spectra of DCN ($3-2$) and C$^{18}$O ($2-1$). The comparison shows that more than 90\% is filtered out in the PRODIGE C$^{18}$O ($2-1$) NOEMA data compared to the merged ANTIHEROES-PRODIGE data. Most transitions (CO, C$^{18}$O, SO, DCN, c-C$_{3}$HC$_{2}$) studied in this work show extended emission and thus including short spacing data is crucial to properly analyze the kinematic properties and measure reliable fluxes. 
	
\section{Results}\label{sec:results}

	Here we present our analysis of the standalone single-dish data, as well as the combined NOEMA + IRAM\,30m molecular line data. In Sect. \ref{sec:kinematics} we describe how the complex line profiles of C$^{18}$O and SO$_{2}$ are disentangled in order to extract infalling streamers as well as connect their kinematic properties to the larger scale environment. In Sect. \ref{sec:radtrans} we use c-C$_{3}$H$_{2}$ to estimate the temperature in the L1448N region. In Sect. \ref{sec:pvdiagram} we infer the source velocity and central mass of the IRS3C system. A streamline model is applied in Sect. \ref{sec:streamline} to the C$^{18}$O and SO$_{2}$ streamers found towards IRS3C. The mass and infall rates of the C$^{18}$O and SO$_{2}$ IRS3C and DCN streamers of IRS3A \citep{Gieser2024} are inferred in Sect. \ref{sec:infallrates}.

\begin{figure*}[!htb]
\centering
\includegraphics[width=0.8\textwidth]{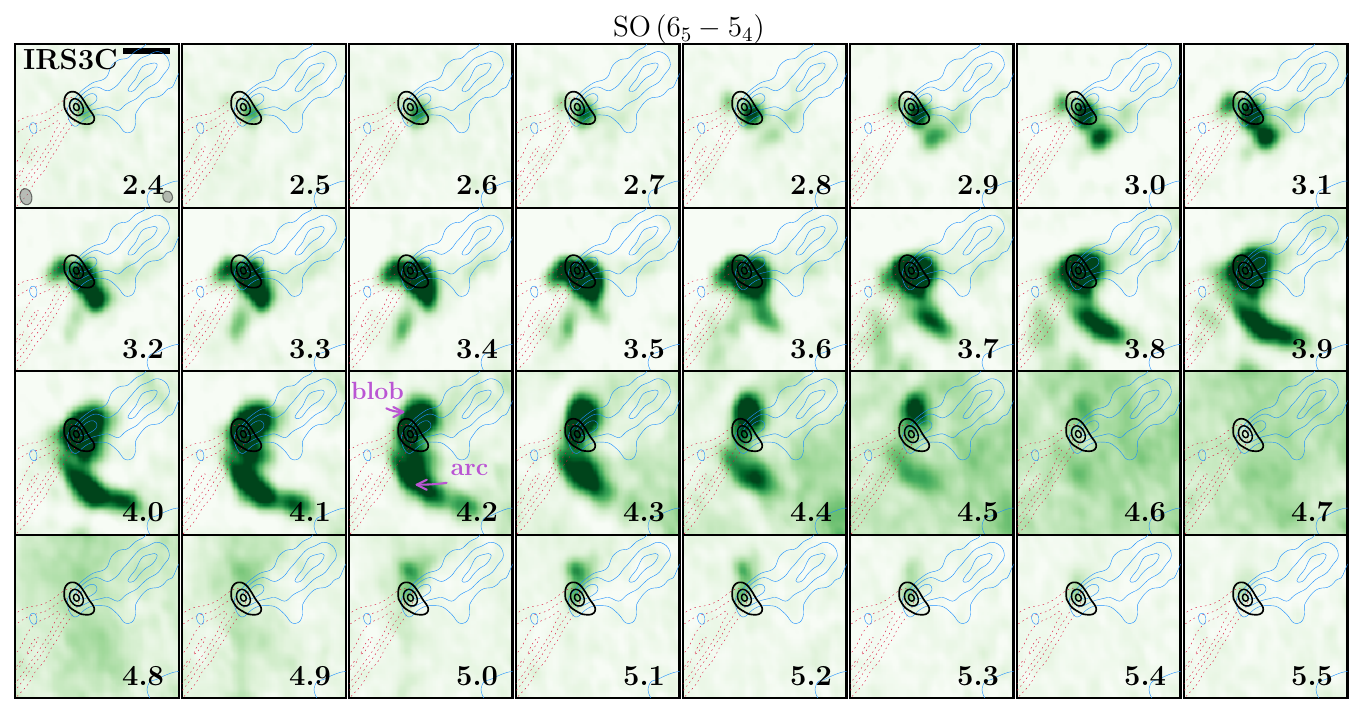}
\includegraphics[width=0.8\textwidth]{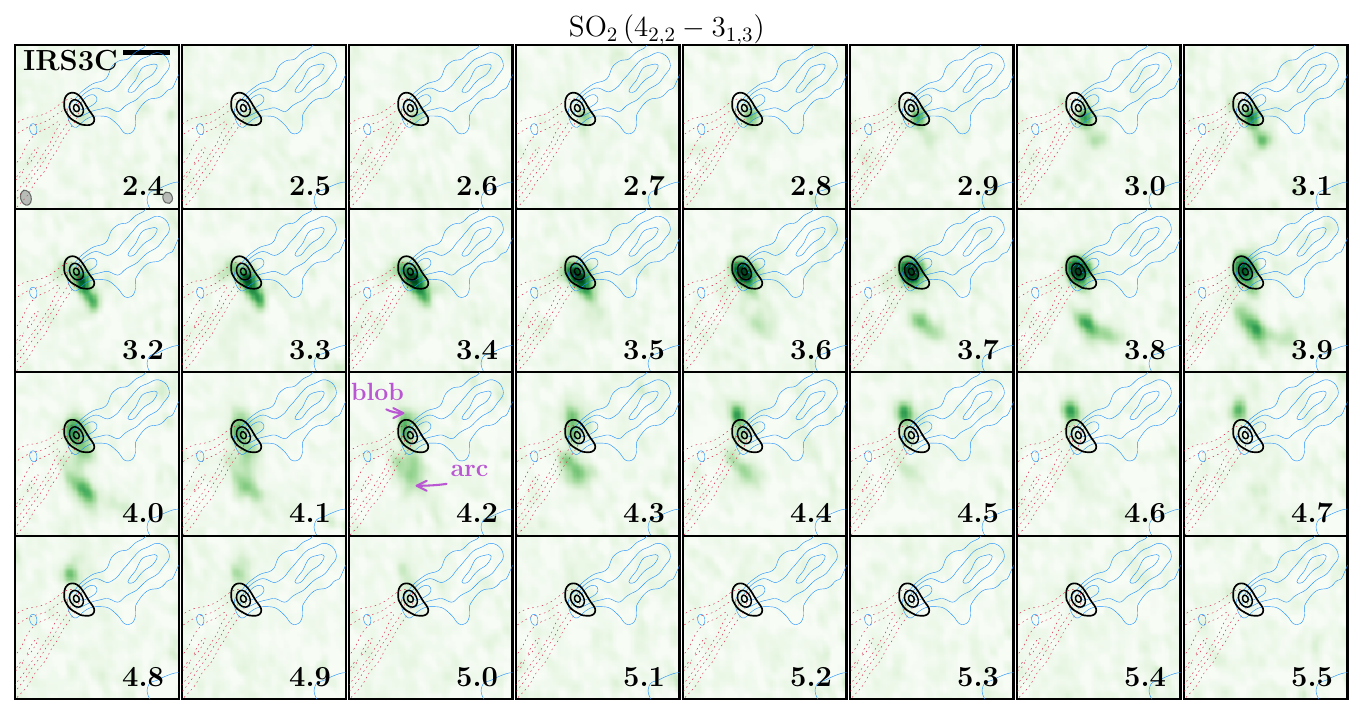}
\caption{Channel maps of SO 6$_5-5_4$ (top, merged data) and SO$_{2}$ $4_{2,2}-3_{1,3}$ (bottom, NOEMA-only data) towards L1448 IRS3C. The green color map shows the line intensity with the velocity of the channel in km\,s$^{-1}$ labeled in the bottom right. The black contours are the same as in Fig. \ref{fig:continuum}. The red dashed and blue solid contours indicate the CO ($2-1$) integrated intensity of the red and blueshifted outflow launched by IRS3C. The integration ranges are 5.9$-$27\,km\,s$^{-1}$ and 2$-$3.4\,km\,s$^{-1}$ (narrow range to avoid foreground contamination) with contour levels at 0.1, 0.3, 0.5, 0.7 and 0.5, 0.7, 0.9 $\times$ peak integrated intensity, respectively. In the top left panel, the black scale bar in the top right indicates a spatial scale of 1\,000\,au and the synthesized beam of the line and continuum data are shown in the bottom left and right, respectively. In each panel the scale is set from a minimum brightness temperature of $-$0.3\,K to a maximum of 7\,K. The IRS3C source velocity is 3.8\,km\,$^{-1}$ (Sect. \ref{sec:pvdiagram}).}
\label{fig:channel_map}
\end{figure*}

\subsection{Kinematic properties of the molecular gas}\label{sec:kinematics}

\subsubsection{Spatial morphology of C$^{18}$O}

	The line integrated ($\varv$=1$-$11\,km\,s$^{-1}$) intensity maps of the C$^{18}$O ($2-1$) transition of the IRAM\,30m and merged data are presented in Fig. \ref{fig:C18OSOSO2mom0}. In the IRAM\,30m map, C$^{18}$O is brightest towards the locations of the three protostellar systems with a NW-SE elongation. In the merged data, this elongated feature shows even more substructure. The bridge structure discovered previously in the PRODIGE data embedding IRS3A and IRS3B is clearly detected \citep{Gieser2024}.
	
	North of IRS3A there is an elongated feature, referred to as the stick in the following. This structure is nearby the IRS3-ALMA source recently detected in mm continuum by \citet{Looney2025}. Although these authors argue that this source is most likely a background galaxy, the stick structure could be associated with the source given the spatial vicinity. IRS3-ALMA could therefore also be a very faint and young protostar or prestellar core associated with L1448N. Further investigating the nature of IRS3-ALMA is beyond the scope of this work.
	
	Surrounding the IRS3C system there are elongated structures. The feature connecting IRS3A/B to IRS3C has previously been reported by \citet{Volgenau2006} in C$^{18}$O ($1-0$) emission (their Fig. 1). Considering the orientation along the outflow direction of IRS3C, the morphology indicates that C$^{18}$O is tracing the remnant envelope carved by the outflow.

\subsubsection{Spatial morphology and channel maps of SO and SO$_2$}
	
	The line integrated ($\varv$=2$-$6\,km\,s$^{-1}$) intensity maps of the SO ($6_{5}-5_{4}$) and SO$_{2}$ ($4_{2,2}-3_{1,3}$) transitions of the IRAM\,30m and merged and NOEMA data, respectively, are presented in Fig. \ref{fig:C18OSOSO2mom0}. In the single-dish data, SO$_{2}$ is not detected, while SO shows a large-scale distribution peaking towards IRS3C but also towards the east and west of the L1448N region where no known protostars are located. In the NOEMA data however, SO$_{2}$ is bright towards IRS3C with a bright arc towards the south of the protostellar system. This elongated arc structure is also detected in bright SO emission.
	
	Channel maps from 2.4\,km\,s$^{-1}$ to 5.5\,km\,s$^{-1}$ towards IRS3C for both the SO ($6_{5}-5_{4}$) and SO$_{2}$ ($4_{2,2}-3_{1,3}$) transitions are presented in Fig. \ref{fig:channel_map}. For comparison, the red and blueshifted morphology of the outflow is highlighted using CO ($2-1$) merged data of ANTIHEROES-PRODIGE (Table \ref{tab:obs}). There is compact SO and SO$_{2}$ emission at high velocities towards the IRS3C protostellar system with rotation from the southwest (blueshifted) to the northeast (redshifted), consistent with high angular resolution C$^{18}$O data tracing the circumbinary material \citep{Tobin2018}. However, the redshifted side is heavily blended by extended emission: In both tracers, from 3.6\,km\,s$^{-1}$ to 4.6\,km\,s$^{-1}$ there is a bright blob towards the north of the IRS3C continuum and the elongated arc structure towards the south-west (marked in the 4.2\,km\,s$^{-1}$ channel in Fig. \ref{fig:channel_map}). In the SO channel map, there is in addition extended SO emission close to the region with velocities between 4.3\,km\,s$^{-1}$ and 4.8\,km\,s$^{-1}$. The emission of the SO blob disappears at 4.7\,km\,s$^{-1}$ and 4.8\,km\,s$^{-1}$, which might be due to absorption against the extended component, as it is still detected in the SO$_{2}$ channel maps. At larger velocities, the blob becomes bright again. Regarding the extended emission of SO, it is known to be abundant in the core envelope, as it has also been detected in starless and prestellar cores \citep{Swade1989,Spezzano2017,Vastel2018,Harju2020,Lattanzi2020,RodriguezBaras2021}. Although SO is brighter and morphologically more complex, we perform the following kinematic analysis on SO$_{2}$ because it does not have additional extended emission.

\subsubsection{Gaussian decomposition and clustering}

	In the following, we analyze the kinematic properties of the molecular gas that has a complex spatial morphology (Fig. \ref{fig:C18OSOSO2mom0}) both in C$^{18}$O and in sulfur-bearing species (SO and SO$_2$). We aim to study in more detail the nature of asymmetric infalling streamers and the connection to the surrounding envelope material. The dynamics of the infalling gas bridge towards IRS3A have been analyzed in detail using DCN ($3-2$) emission in \citet{Gieser2024}, see also Fig. \ref{fig:comparison}. While the previous NOEMA-only data could not be used to analyze brighter and more extended molecular emission, such as C$^{18}$O, the complementary ANTIHEROES data allows a detailed kinematic analysis over a large area. Here, we apply the same method presented in \citet{Gieser2024} to the C$^{18}$O ($2-1$) single-dish and merged data, as well as the SO$_{2}$ ($4_{2,2}-3_{1,3}$) NOEMA data to infer the large-scale kinematic properties of the L1448N envelope as well as the smaller-scale properties towards IRS3C.

	To obtain the kinematic properties, we fit up to three Gaussian velocity components with the \texttt{pyspeckit} package \citep{Ginsburg2022} to each position where the peak S/N in that spectrum is larger than 5. For further details of the methodology we refer to \citet{Gieser2024} and a summary is presented in Appendix \ref{app:kinematic}.

\begin{figure*}[!htb]
\centering
\includegraphics[width=0.37\textwidth]{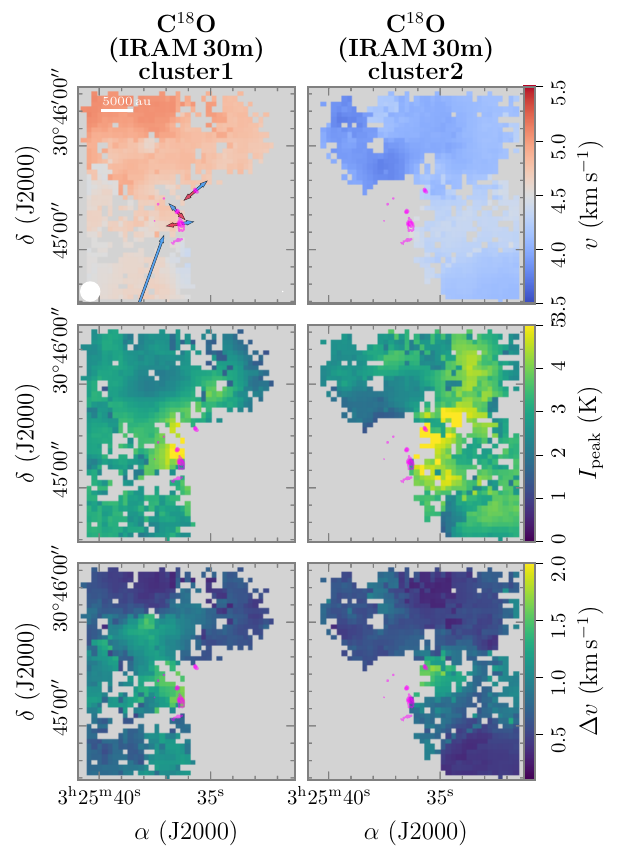}
\includegraphics[width=0.41\textwidth]{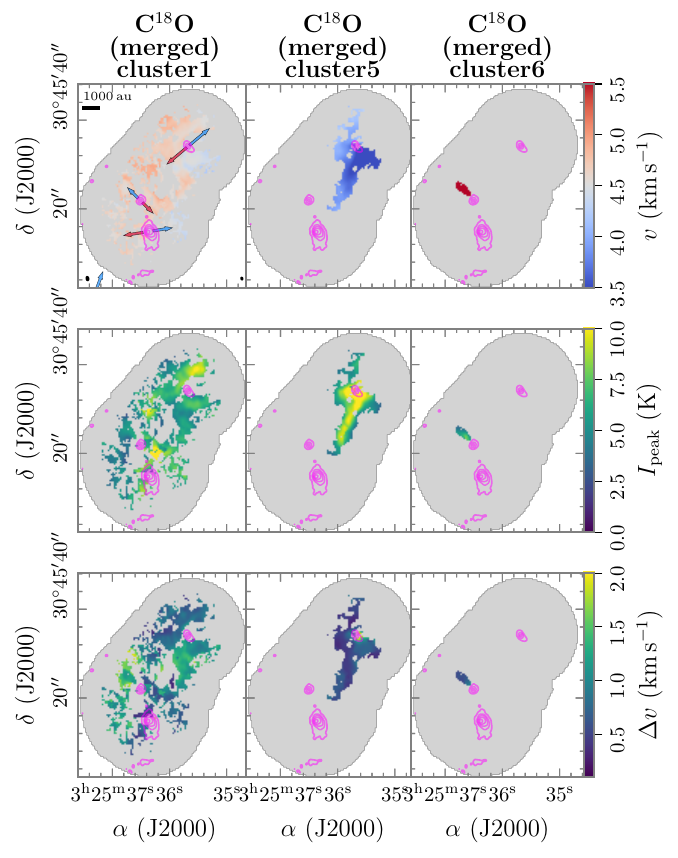}
\includegraphics[width=0.205\textwidth]{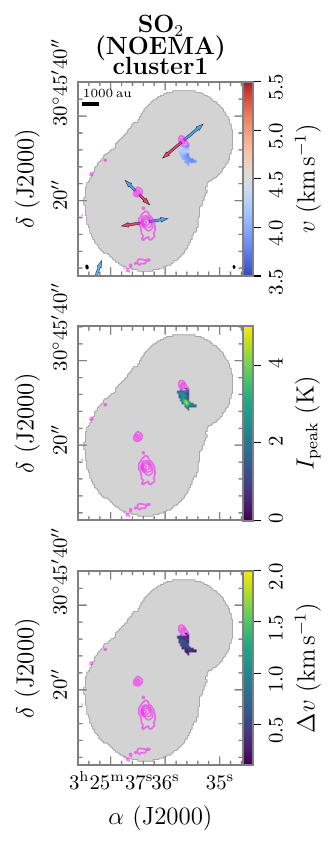}
\caption{Main clusters extracted with \texttt{DBSCAN}. The velocity, peak intensity, and line width are shown in color in the top, central, and bottom panels, respectively. Contours and arrows are the NOEMA 1.4\,mm continuum with the same levels as in Fig. \ref{fig:continuum}. In cluster 1 of each analyzed data set, the synthesized beam of the line and continuum data is shown in the bottom left and right corner, respectively, and scale bars are shown in the top left corner. The remaining clusters are shown in Fig. \ref{fig:clusters_app}.}
\label{fig:clusters}
\end{figure*}

	Following the methodology by \citet{Gieser2024}, we use the Density-based spatial clustering of applications with noise (\texttt{DBSCAN}) tool of the \texttt{scikit-learn} python package \citep{scikit-learn} to extract individual cluster from the data set of extracted velocity components. The input is the position (x,y) as well as peak velocity $\varv$ of all velocity components after quality assessment (Appendix \ref{app:kinematic}). We note that the line width and amplitude are not used as an input, but are used to confirm that the clustering results provide reliable results not only in the velocity, but also in line width and amplitude maps. The main clusters relevant for this work are presented in Fig. \ref{fig:clusters}, while the remaining clusters as well as data points discarded by \texttt{DBSCAN} are shown in Fig. \ref{fig:clusters_app} in the Appendix.
	
	On large scales as traced by the IRAM\,30m data, C$^{18}$O has a velocity gradient from east (red-shifted, C$^{18}$O (IRAM\,30m) cluster 1) to west (blueshifted, C$^{18}$O (IRAM\,30m) cluster 2) indicating rotation of the large scale envelope material (left panel in Fig. \ref{fig:clusters}). The protostellar systems are oriented along the rotation axis of this large scale motion, i.e. they are located where the velocities change from red- to blue-shifted. This large-scale motion is consistent with the rotation of the disks \citep{Tobin2018, Reynolds2021}. The infalling DCN bridge also shows the same velocity pattern with the northern redshifted streamer on the east and the southern blueshifted streamer on the west \citep{Gieser2024}. The agreement with the large-scale C$^{18}$O velocity map thus implies that the infalling material from the gas bridge originates from the large-scale environment. We further discuss this result in Sect. \ref{sec:disc_kin}. Towards the protostellar systems there is in addition an increase of line width with $\Delta \varv$ up to 1.5\,km\,s$^{-1}$, while in the surrounding environment $\Delta \varv <$0.5\,km\,s$^{-1}$. Such an increase in line width along the streamer towards the protostellar system is found towards other streamers such as IRS3A \citep{Gieser2024} and Per-emb-50 \citep{ValdiviaMena2022}.
	
	The large-scale envelope rotation is also seen in the higher angular resolution C$^{18}$O data set (C$^{18}$O (merged) cluster 1). The redshifted side of the infalling bridge towards IRS3A can also be extracted in the merged C$^{18}$O data set (C$^{18}$O (merged) cluster 6). We note that due to the large amount of data points and complex line profiles close to the infalling bridge, it is not possible to extract the blue side of the IRS3A bridge in C$^{18}$O, even though it is evident in the integrated intensity map (Fig. \ref{fig:C18OSOSO2mom0}). In addition to the rotation of the envelope, there is an extended emission feature surrounding IRS3C (C$^{18}$O (merged) cluster 5), where velocities become more blueshifted as the system is approached from the south. The DCN bridge around IRS3A showed a very similar behavior, i.e. velocities become more red- and blue-shifted closer to the protostar, which is evidence for infall motions.
	
	The NOEMA SO$_{2}$ clustering results extract the arc structure towards IRS3C (SO$_{2}$ (NOEMA) cluster 1) where the emission becomes more redshifted as the system is approached. In higher resolution ALMA data, \citet{ArturdelaVillarmois2023} detect this extended arc in SO and SO$_{2}$ emission (with transitions at higher upper energy levels of $48-95$\,K). This suggests that along the arc itself the gas is denser and/or hotter compared to the rest of the gas. The C$^{18}$O and SO$_{2}$ asymmetric structures (C$^{18}$O (merged) cluster 5 and SO$_{2}$ (NOEMA) cluster 1) extracted from this clustering method and their infall motions onto IRS3C will be further characterized in Sect. \ref{sec:streamline}.
	
\subsection{Temperature distribution in L1448N derived from c-C$_{3}$H$_{2}$}\label{sec:radtrans}

\begin{figure}[!htb]
\centering
\includegraphics[width=0.49\textwidth]{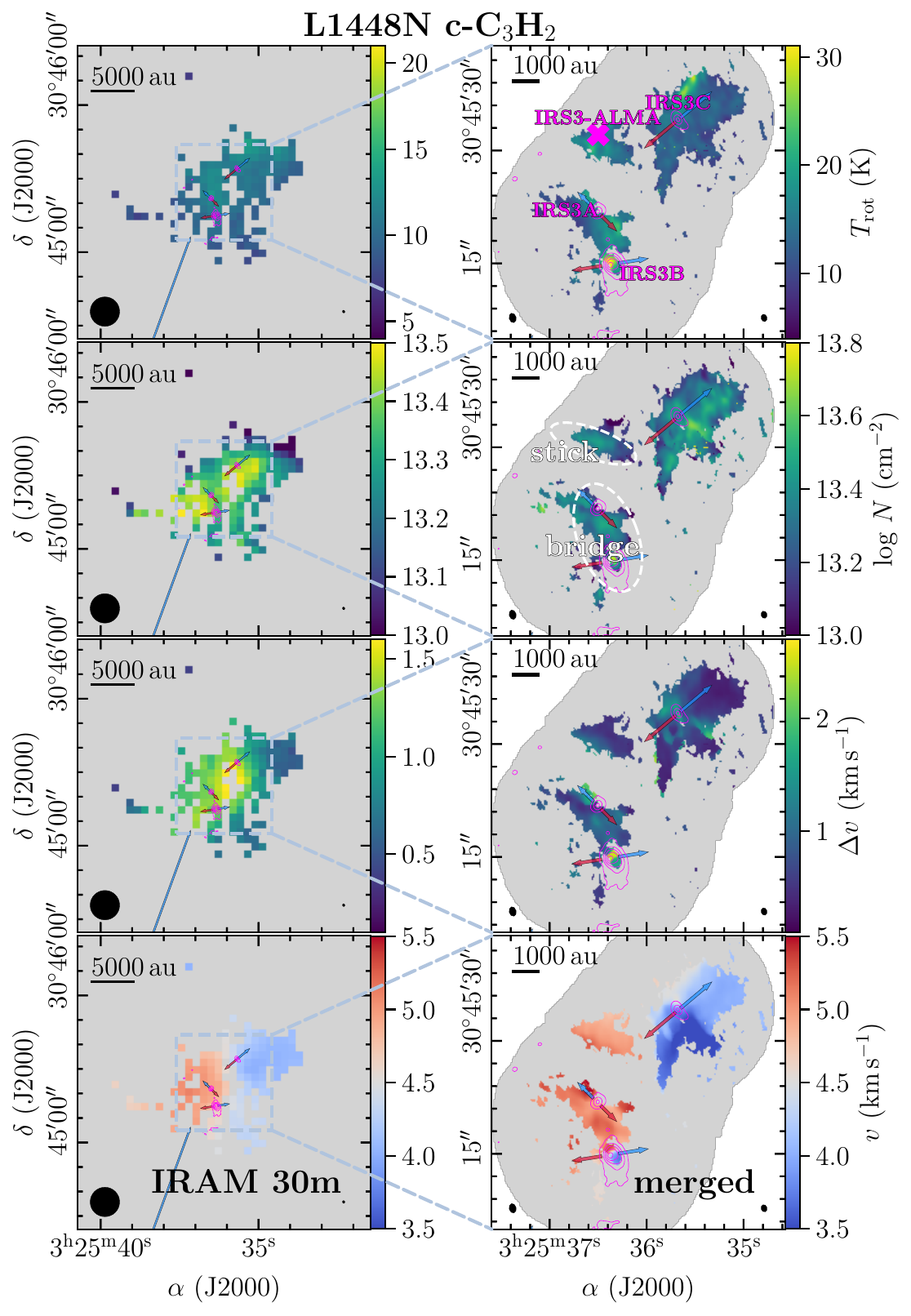}
\caption{Parameters maps of c-C$_{3}$H$_{2}$ (rotation temperature, column density, FWHM line width, and peak velocity) of the IRAM\,30m (left) and merged (right) data. Contours and arrows are the same as in Fig. \ref{fig:continuum}. The beam of the line and continuum data are shown in the bottom left and right corner, respectively. Scale bars are indicated in the top left corner.}
\label{fig:xclass_C3H2}
\end{figure}	

	While hot corino sources in the PRODIGE sample show a plethora of emission lines by complex organic molecules \citep[][]{Hsieh2024,Busch2025}, the PRODIGE spectral setup typically covers only one transition of smaller molecules. One of the exceptions is c-C$_{3}$H$_{2}$ for which three transitions are detected in the PRODIGE+30m data in the L1448N region (Table \ref{tab:obs}). The velocity integrated intensity maps (integrated between 3 and 7\,km\,s$^{-1}$) of the IRAM\,30m and merged data are presented in Fig. \ref{fig:mom0_C3H2}.
	
	
	
	The emission of the three c-C$_{3}$H$_{2}$ transitions (Fig. \ref{fig:mom0_C3H2}) is spatially extended in the 30m data, embedding the three protostellar systems, similar to the distribution of C$^{18}$O. In the high resolution NOEMA+30m data, the gas bridge, the stick, as well as the elongated envelope surrounding IRS3C are traced. The extended emission of c-C$_{3}$H$_{2}$ suggests that L1448N is a warm carbon chain chemistry (WCCC) source (even though c-C$_{3}$H$_{2}$ is the cyclic isomer), as suggested previously by \citet{Sakai2009}. The authors found a high abundance ($>$10$^{-9}$) of C$_4$H derived from single-dish observations and proposed L1448N as a candidate WCCC source.

	The underlying excitation conditions, such as rotation temperature and column density can be inferred by fitting the radiative transfer equation of the rotational transitions. We use the eXtended CASA Line Analysis Software Suite \citep[\texttt{XCLASS},][version 1.4.3]{XCLASS} to fit the observed IRAM\,30m and merged spectral line data of c-C$_{3}$H$_{2}$. For both data sets, we apply the same methodology as explained in detail in Appendix \ref{sec:app_linemodeling}. Although spectral line profiles of c-C$_{3}$H$_{2}$ show multiple components in some locations (e.g. close to the protostars, Fig. \ref{fig:spectra}), we choose to fit the observed line profiles with only one velocity component in order to obtain the average excitation conditions in the region. The resulting best-fit parameter maps ($T_\mathrm{rot}$, $N$, $\Delta \varv$, $\varv$) of c-C$_{3}$H$_{2}$ are presented in Fig. \ref{fig:xclass_C3H2}.
	
	The mean excitation temperature in the region is 10\,K and 14\,K in the 30m and merged data, respectively, indicating that over a large area, the molecular gas is still cold in the L1448N region. The merged data however shows areas with enhanced temperatures, for example, towards the IRS3B disk with temperatures higher than 30\,K. The triple system embedded in the disk is efficiently heating up the surrounding gas. The northern envelope material of IRS3C also has elevated temperatures (25\,K) which is gas being heated up along the cavity walls of the outflow. 
	
	In both data sets the mean column density is similar, at $\approx$2$\times$10$^{13}$\,cm$^{-2}$, but the merged data is spatially resolving higher column density regions towards the protostellar systems as well as the surrounding envelope material. Notably, the stick is located very close to IRS3-ALMA (a projected distance of $\approx1''$ corresponding to $\approx$300\,au) and also shows elevated temperatures and column densities in the merged data. Despite modeling only one c-C$_{3}$H$_{2}$ emission component, the line width and velocity maps are similar to those extracted from the more sophisticated kinematic analysis of C$^{18}$O in Sect. \ref{sec:kinematics}. Thus, C$^{18}$O and c-C$_{3}$H$_{2}$ are tracing the same molecular gas components (a spectrum comparing the line profile of molecular analyzed in this work towards all three protostellar systems is shown in Fig. \ref{fig:spectra}). In Sect. \ref{sec:infallrates} we use the temperature determined by the \texttt{XCLASS} fitting of c-C$_{3}$H$_{2}$ in order to estimate the mass of infalling streamers and the L1448N envelope.
	
\subsection{Systemic velocity and mass of IRS3C} \label{sec:pvdiagram}

\begin{figure}[!htb]
\centering
\includegraphics[width=0.4\textwidth]{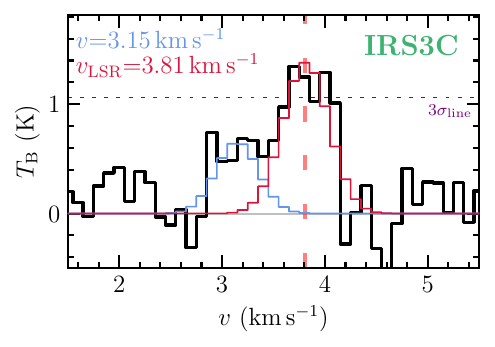}
\caption{Spectrum of DCN ($3-2$) extracted from the IRS3C continuum peak position. The observed spectrum is shown in black and a two Gaussian velocity component fit is shown in blue and red. The purple dashed line marks the $3\sigma_\mathrm{line}$ level. The source velocity estimated from the red component is indicated by a vertical dashed line.}
\label{fig:sourcevelocity}
\end{figure}

\begin{figure}[!htb]
\centering
\includegraphics[width=0.4\textwidth]{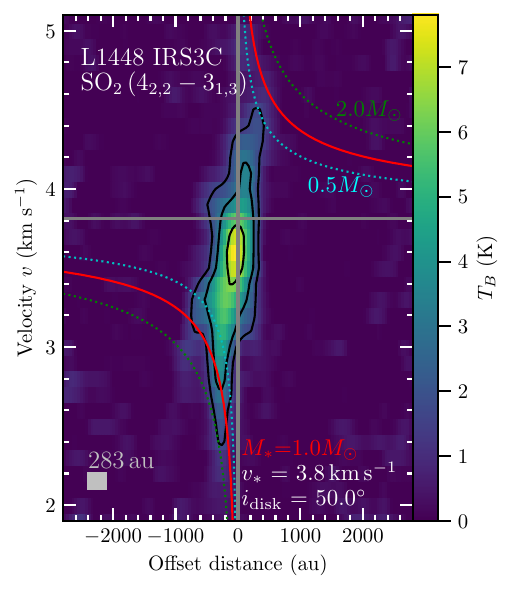}
\caption{Position velocity diagram of SO$_{2}$ ($4_{2,2}-3_{1,3}$) extracted along the disk orientation of IRS3C. Black contour levels are the 4, 8, and 16$\sigma_\mathrm{line}$ levels. The red line shows the profile for a Keplerian disk of 1\,$M_\odot$ inclined by 50$^\circ$ and for comparison the cyan and green dotted lines show profiles for 0.5\,$M_\odot$ and 2\,$M_\odot$, respectively. A spectral and spatial resolution element (0.1\,km\,s$^{-1} \times$283\,au) is indicated in the bottom left.}

\label{fig:pvdiagram}
\end{figure}

	In order to test if the structures detected towards IRS3C in C$^{18}$O and SO$_{2}$ (Fig. \ref{fig:clusters}) are due to gravitational infall using streamline models \citep[][Sect. \ref{sec:streamline}]{Pineda2020}, the precise source velocity as well as the central mass of IRS3C have to be determined first. We estimate the source velocity using DCN ($3-2$) emission following the approach by \citet{Gieser2024}. DCN only has a weak detection towards the continuum peak position of IRS3C (Fig. \ref{fig:sourcevelocity}), but the spectral line profile is not as complex as other species such as C$^{18}$O and SO$_{2}$ (Fig. \ref{fig:spectra}). We fit two Gaussian velocity components to the DCN ($3-2$) spectral profile and assign the brighter component to the IRS3C source velocity, at $\varv_\mathrm{LSR}=3.81\pm0.05$\,km\,s$^{-1}$. Despite its low S/N, the second component is included in the fit to avoid biasing the fitted velocity of the brighter component. The peak of the brighter component coincides with the peak emission of other molecular transitions (Fig. \ref{fig:spectra}) and is hence a good estimate of the IRS3C source velocity.
	
	The central mass can be estimated from fitting the position velocity (PV) diagram of molecular lines tracing the disk rotation. In the case of IRS3A the C$^{18}$O ($2-1$) transition was used \citep{Gieser2024}, however, around the IRS3C system the emission is too complex with a significant distribution of the surrounding envelope material (Fig. \ref{fig:C18OSOSO2mom0}). Here, we use the SO$_{2}$ ($4_{2,2}-3_{1,3}$) transition to estimate the central mass, however, since it is faint, we are not able to fit the PV data (Fig. \ref{fig:pvdiagram}), but matching the observations by eye can still provide an estimate of the mass \citep[see also, e.g., ][]{Reynolds2021,ValdiviaMena2022}. Given the low S/N of the SO$_{2}$ PV diagram, more sophisticated PV diagram modeling including rotation and infall are needed, but are beyond the scope of this work. Extracting the PV diagram using the \texttt{pvextractor} package \citep{Ginsburg2015}, the blueshifted SO$_{2}$ velocities appears roughly compatible with an inclination-corrected mass of $\approx$1\,$M_\odot$ (Fig. \ref{fig:pvdiagram}). The position angle and inclination of the disk were taken from \citet{Reynolds2024}. For comparison, we also show Keplerian models for inclination corrected masses of 0.5\,$M_\odot$ and 2\,$M_\odot$.

\subsection{Streamline model}\label{sec:streamline}

\begin{figure*}[!htb]
\centering
\includegraphics[width=0.4\textwidth]{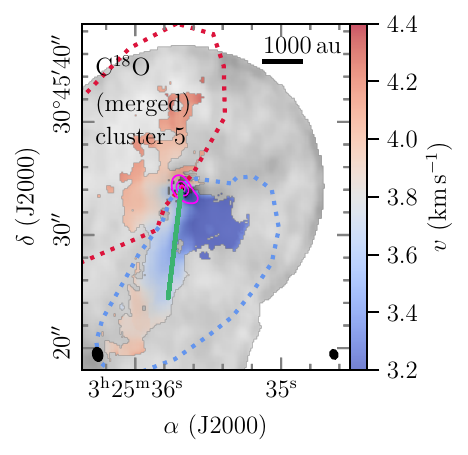}
\includegraphics[width=0.4\textwidth]{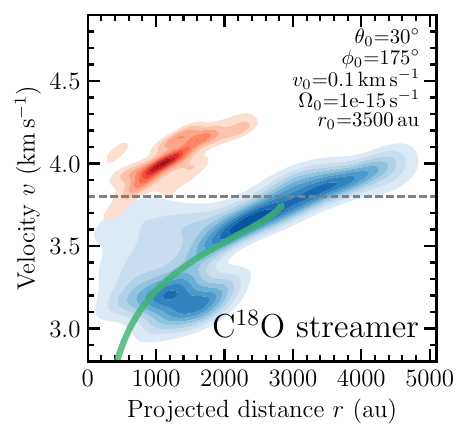}\\
\includegraphics[width=0.4\textwidth]{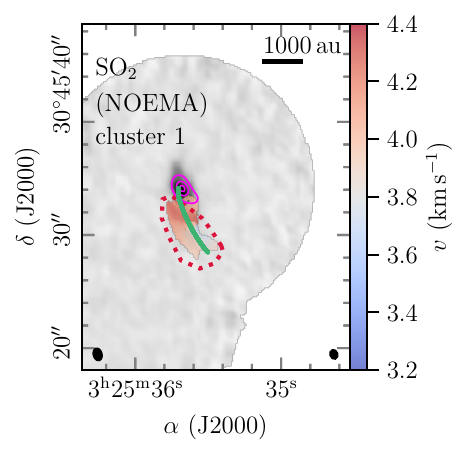}
\includegraphics[width=0.4\textwidth]{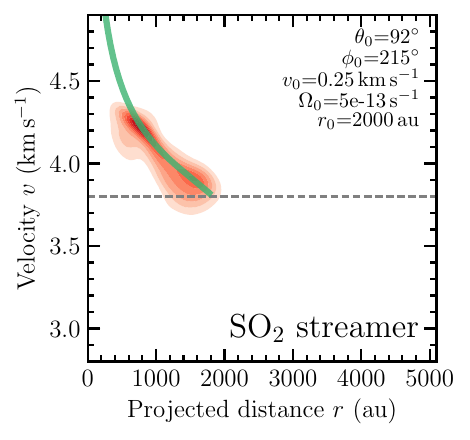}
\caption{Streamline models towards IRS3C. In the left panels, the line integrated intensity and the velocities of the structure are shown in grey and color, respectively. The trajectory of the best match streamline model is shown in green. Contours are the NOEMA 1.4\,mm continuum with the same levels as in Fig. \ref{fig:continuum}. The beam of the line and continuum data are shown in the bottom left and right corner, respectively. A scale bar is shown in the top right corner. In the right panels the velocity distribution extracted from the areas indicated by dotted ellipses on the left is shown in red or blue color and the best match streamline model is shown in green. The source velocity of IRS3C (3.8\,km\,s$^{-1}$) is indicated by the horizontal dashed line.}
\label{fig:streamers}
\end{figure*}

	Combining the observed velocity profiles of the C$^{18}$O and SO$_2$ asymmetric structures (C$^{18}$O (merged) cluster 5 and SO$_{2}$ (NOEMA) cluster 1) in Fig. \ref{fig:clusters}) that show signs of infall motions towards IRS3C, as well as the determined source velocity and central mass (Sect. \ref{sec:pvdiagram}), it is now possible to test if the velocities are consistent with gravitational infall towards IRS3C using a streamline model. The analytical solutions \citep{Mendoza2009} and the applications to infalling streamers are described in detail in \citet{Pineda2020} and are implemented in the \texttt{velocity\_tools} python package\footnote{\url{https://github.com/jpinedaf/velocity_tools}}. In short, the streamline model follows the trajectory of a particle in the rotating envelope and is infalling towards the central gravitational sink, i.e., the protostellar system.
	
	We run a grid of models varying the radial separation $r_0$, initial position ($\theta_0$, $\phi_0$), radial velocity $\varv_0=\varv$($r_0$) and angular velocity ($\Omega_0$) of the particle. We set the rotation of the system to the observed rotation of the circumbinary envelope \citep{Tobin2018}. We note that the angular resolution is not sufficient to spatially resolve the disks of each binary, so we treat the system as one. This also implies that we can not determine if the streamers feed only one of the protostars in the binary or both. 
	
	The observed velocity profiles as a function of projected distance from the protostar are shown in the right in Fig. \ref{fig:streamers}. For the blue side of the C$^{18}$O structure and the SO$_{2}$ structure the velocities become more blueshifted and redshifted, respectively, as the protostars are approached. For the redshifted side of the C$^{18}$O structure however, we find that the velocity is becoming more redshifted with distance from the protostar, opposite of what is expected from an infalling structure. However, since only the line-of-sight velocity component can be observed, it could still be infalling. Figure \ref{fig:streamers} shows the best match trajectory projected along the plane of the sky on the left as well as the observed and streamline velocity profile for the C$^{18}$O and SO$_{2}$ streamers on the right.
	
	For the blue C$^{18}$O and SO$_{2}$ infalling structures we find a very good match in both the trajectory as well as velocity profile in the streamline model. For C$^{18}$O we do not find a cluster that matches the SO$_{2}$ streamer velocity profile, which could be due to the high spectral complexity of C$^{18}$O line profiles towards the protostellar systems (Fig. \ref{fig:spectra}). IRS3C is the first example that shows asymmetric infall from the remnant envelope (C$^{18}$O) with an additional streamer (bright in SO and SO$_{2}$) that cannot be directly connected to the envelope material.

\subsection{Streamer mass and infall rates}\label{sec:infallrates}

		In \citet{Gieser2024} both the red and blue side of the DCN bridge show infall motions towards IRS3A. However, since at the time short spacing data were not yet available, mass infall rates were not determined. A comparison of the NOEMA-only and merged DCN data is presented in Appendix \ref{sec:app_merging}. The DCN bridge surrounding IRS3A is not affected by spatial filtering and thus in the following we use the kinematic results by \citet{Gieser2024} to estimate the mass and infall rates of the streamers towards IRS3A. For IRS3C we estimate the mass and infall rates of the C$^{18}$O and SO$_{2}$ streamers found in this work (Fig. \ref{fig:streamers}).
	
	Molecular column density ($N$) maps of the C$^{18}$O and DCN streamers in IRS3C and IRS3A, respectively, are calculated using an optically thin approximation \citep[Eq. 80 in][]{Mangum2015} within the \texttt{molecular\_columns} python package\footnote{\url{https://github.com/jpinedaf/molecular_columns}}. For the excitation temperature we use $T_\mathrm{rot}$ from the c-C$_{3}$H$_{2}$ map, assuming $T=10$\,K in all pixels with no available c-C$_{3}$H$_{2}$ temperature estimate. The integrated intensity is calculated using the Gaussian fit results (intensity and line width) of the corresponding cluster. The abundance ratio $X$ of C$^{18}$O/H$_{2}$ and DCN/H$_{2}$ are assumed to be $1.8\times10^{-7}$ \citep{Frerking1982,Wilson1994} and $2.4\times10^{-10}$ \citep{Hsieh2023}, respectively. To compare the streamer masses with the total envelope mass traced by the IRAM\,30m C$^{18}$O data, we apply the same method of column density calculation using the C$^{18}$O integrated intensity map (Fig. \ref{fig:C18OSOSO2mom0}) and the temperature from the c-C$_{3}$H$_{2}$ single-dish map assuming $T=10$\,K in all pixels with no available c-C$_{3}$H$_{2}$ temperature estimate.
	
	A column density map of the SO$_{2}$ streamer is created using \texttt{XCLASS} (a description of the tool is presented in Sect. \ref{sec:radtrans}). We fix the source size and rotation temperature parameters to $\theta_\mathrm{source}=2''$ and $T_\mathrm{rot}=T_\mathrm{rot}$(c-C$_{3}$H$_{2}$) and $T_\mathrm{rot}$=10\,K otherwise). The line width and velocity are also fixed according to the results from the clustering (SO$_{2}$ (NOEMA) cluster 1, Fig. \ref{fig:clusters}). The column density is then the remaining free parameter. The abundance ratio $X$ of SO$_{2}$/H$_{2}$ is assumed to be $10^{-10}$ \citep{ArturdelaVillarmois2023,MiranzoPastor2025}. The column density maps of the streamers as well as the large-scale envelope are presented in Fig. \ref{fig:coldens}. 
	
		The total mass per pixel is calculated as $N \times (\Delta s)^2 \times m(H_2) / X$, with pixel scale $\Delta s$. The total mass of the streamers ($M_\mathrm{streamer}$) and envelope ($M_\mathrm{env}$) is the sum of all pixels. The free-fall timescale is calculated as $t_\mathrm{ff} = \sqrt{R_\mathrm{streamer}^3/(G \times M_\mathrm{*})}$ with streamer length $R_\mathrm{streamer} = r_0$ determined from the streamline model (Sect. \ref{sec:streamline}). Here we neglect the contribution of the initial velocity $\varv_0$ of the best match streamline model. The mass $M_\mathrm{*}$ of the central protostellar system was estimated from the PV diagram of C$^{18}$O and SO$_{2}$ for IRS3A \citep{Gieser2024} and IRS3C (this work, Sect. \ref{sec:pvdiagram}), respectively. The mass infall rate $\dot M_\mathrm{streamer}$ is $\dot M_\mathrm{streamer} = M_\mathrm{streamer} / t_\mathrm{ff}$. The total mass of the envelope is 16.4\,M$_\odot$ consistent with recent estimates from dust continuum emission \citep[$\approx$15.9\,M$_\odot$][]{Murillo2024}. The streamer properties are summarized in Table \ref{tab:infall} and are further discussed in the next section. 
		
		The uncertainties of the streamer and envelope mass, as well as infall rate calculation are dominated by the uncertainties of the abundance ratios. Here we use average values at the Galactic distance of the Perseus star-forming region for the C$^{18}$O/H$_{2}$ ratio. For DCN/H$_{2}$ we use estimates from the nearby SVS13A protostar which is also located in the same molecular cloud. However, local chemical differences could cause a significant deviation from these values. In the case of SO$_{2}$/H$_{2}$, \citet{ArturdelaVillarmois2023} measured a ratio of $1.4\pm0.5\times10^{-10}$ towards L1448 IRS3C within an aperture with a diameter of 0.64$''$ consistent with recent estimates from PRODIGE towards the source \citep[1.3$\times10^{-11}$-1.2$\times10^{-10}$,][]{MiranzoPastor2025}. However, along the streamer, this ratio could also differ. Hence, here we can assume that the streamer and envelope masses and infall rates are uncertain by a factor of a few.
		
\section{Discussion}\label{sec:discussion}


\begin{figure}[!htb]
\centering
\includegraphics[width=0.49\textwidth]{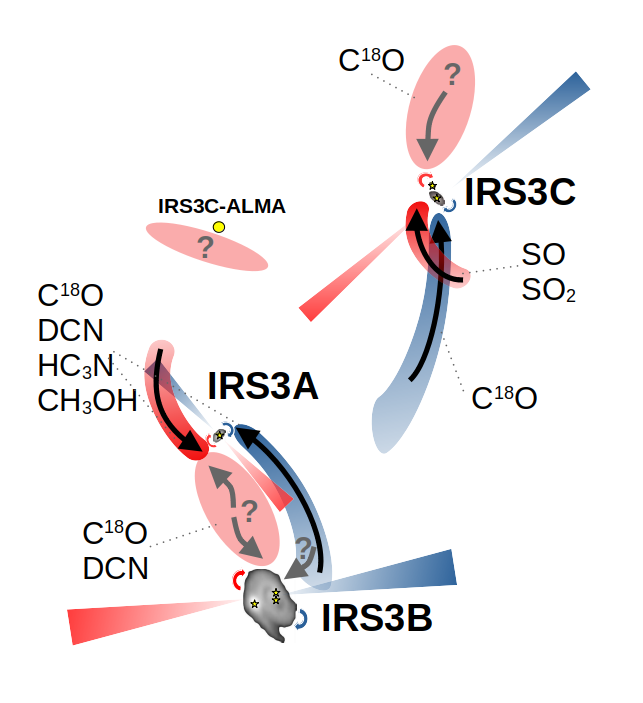}
\caption{Sketch of L1448N \citep[expanded from Fig. 10 in][]{Gieser2024}.}
\label{fig:sketch}
\end{figure}

\setlength{\tabcolsep}{5pt}
\begin{table*}[!htb]
\caption{Source and streamer properties.}
\label{tab:infall}
\centering
\renewcommand{\arraystretch}{1.5}
\begin{tabular}{llllll | lrrr}
\hline\hline
Source & Multiplicity & & & Source velocity & Central mass & \multicolumn{4}{c}{\underline{Infalling streamers}} \\
 & & $T_\mathrm{bol}$ & $L_\mathrm{bol}$ & $\varv_{\mathrm{LSR}}$ & $M_*$ & streamer & $M_\mathrm{streamer}$ & $t_{ff}$ & $\dot M_\mathrm{streamer}$\\
 & & (K) & ($L_\odot$) & (km\,s$^{-1}$) & ($M_\odot$) & & ($M_\odot$) & ($10^{4}$yr) & ($10^{-6}M_\odot$\,yr$^{-1}$)\\
\hline
IRS3A & single$^{(a)}$ & 47$^{(c)}$ & 9.2$^{(c)}$ & 5.26$^{(b)}$ & 1.2$\pm$0.1$^{(b)}$ & DCN (blue) & 0.0085 & 0.92 & 0.92\\
& & & & & & DCN (red) & 0.011 & 0.92 & 1.2\\
IRS3C & binary$^{(a)}$ & 22$^{(c)}$ & 1.4$^{(c)}$ & 3.81 & $\approx$1.0 & C$^{18}$O & 0.083 & 3.3 & 5.0\\
& & & & & & SO$_{2}$ & 0.17 & 1.4 & 18 \\
\hline
\end{tabular}
\tablefoot{References: $^{(a)}$ \citet{Tobin2018}, $^{(b)}$ \citet{Gieser2024}, $^{(c)}$ \citet{Tobin2016}.}
\end{table*}

The L1448N region consisting of three protostellar systems shows multiple infalling streamers towards IRS3A and IRS3C, revealed by several molecules. A sketch of the streamers in comparison to the bipolar outflow directions and rotation of the disks or circumbinary material in the case of IRS3C is presented in Fig. \ref{fig:sketch}. In the following we discuss the properties of the streamers and their connection to the larger scale environment.

	\subsection{Streamer properties}
		
		An overview of the IRS3A and IRS3C properties and the results of $M_\mathrm{streamer}$, $t_\mathrm{ff}$, and $\dot M_\mathrm{streamer}$ are summarized in Table \ref{tab:infall}. The streamers could provide an additional $\approx$1\% and $\approx$8-17\% of mass towards IRS3A and IRS3C, respectively, compared to the central mass $M_\mathrm{*}$ of the systems, assuming that all of the streamer mass was accreted onto the protostar. The mass infall rates are on the order of $10^{-6}$\,$M_\odot$\,yr$^{-1}$ for IRS3A and considerably higher (5-18$\times10^{-6}$\,$M_\odot$\,yr$^{-1}$) for the younger system IRS3C.
		
		While the protostellar mass accretion rate of IRS3A is $5\times10^{-7}$\,$M_\odot$\,yr$^{-1}$ \citep[derived from the luminosity and stellar mass,][]{Reynolds2021}, the infall rates are a factor of two higher. From mid-infrared imaging, the envelope mass infall rate was estimated to be one order of magnitude higher at $10^{-5}$\,$M_\odot$\,yr$^{-1}$ \citep{Ciardi2003}, whereas IRS3B was not detected in the MIR in that work.
		
		 The IRS3C infalling streamer detected in SO$_{2}$ and SO could hint at a moderate shock and elevated temperatures caused by the higher infall rate as the material is infalling from the larger scale environment to the compact envelope and disk system. ALMA observations of SO$_{2}$ and SO transitions with higher upper energy levels also trace the streamer structure \citep[][]{ArturdelaVillarmois2023}. Such a change in chemical composition where sulfur-bearing species become more abundant has been previously found towards the transition zone from the envelope to disk regime towards the IRAS\,043681+2557 protostar \citep{Sakai2014}. Given that in most cases in the literature, streamers have been found serendipitously in various molecular tracers, their chemical composition in comparison to the surrounding material has not been well studied. Given that not all streamers origin from the natal envelope, it is not necessary that they share similar physical and chemical conditions as the envelope. Such an analysis is now possible with a large sample of Class 0/I protostellar systems in the Perseus star-forming region using data of ANTIHEROES-PRODIGE.
		 
		 Infall rates estimated towards other Class 0/I systems are similar, e.g., $10^{-6}$\,$M_\odot$\,yr$^{-1}$ \citep[Per-emb-2 and Per-emb-50,][]{Pineda2020,ValdiviaMena2022}, $6\times10^{-6}$\,$M_\odot$\,yr$^{-1}$ \citep[HH212,][]{Lee2006}, $3-6\times10^{-6}$\,$M_\odot$\,yr$^{-1}$ \citep[L1527,][]{FloresRivera2021}. We note that the mass of the infalling streamers is not high enough to maintain such high infall rates for a long time and thus the infall rate is expected to vary with time. This is in agreement with simulations by \citet{Kuffmeier2019} that have shown that such structures surrounding and feeding protostellar systems exist up to a few 10\,kyr. However, the IRAM\,30m molecular line maps reveal that there is still a large mass reservoir of $\approx16$\,$M_\odot$ surrounding the L1448N protostellar systems that in the future can supply even more material down to the protostellar scales.
		 
		 A similar study comparing the large-scale envelope and streamer material has been recently carried out by \citet{Kido2025}. The Class 0 protostellar system IRAS\,16544-1604 is being fed by three streamers detected in C$^{18}$O emission with masses of 1-4$\times10^{-3}$\,M$_\odot$ and mass infall rates of 1-5$\times10^{-8}$\,M$_\odot$\,yr$^{-1}$. Using a Keplerian rotation model, the central mass is estimated to be 0.14\,M$_\odot$, much lower compared to IRS3A and IRS3C (Table \ref{tab:infall}). The authors estimate a mass of 0.5\,M$_\odot$ in the envelope, a factor of 30 smaller compared to L1448N. The presence of a larger mass reservoir can explain the higher mass infall rates in the L1448N systems compared to the lower-mass IRAS\,16544-1604 protostar. In agreement with \citet{Kido2025}, we find that the streamer masses in L1448N are much lower ($<$1\%) compared to the overall envelope mass (16.4\,M$_\odot$), implying that a single streamer event does not deliver a significant amount of mass from the envelope to the smaller scales. However, a comparison of the streamer mass with the central mass of the protostellar systems (Table \ref{tab:infall}) shows that a single streamer could significantly impact the masses of the protostellar systems. This could have implications on the disk angular momentum and stability, and even the accretion rate onto the protostar.
	
	\subsection{Kinematic properties from envelope to disk scales}\label{sec:disc_kin}
	
	In this work we have analyzed the kinematic properties of molecular gas from $\approx$36\,000\,au scales down to 300\,au. We find that the velocity profile of the two DCN streamers in IRS3A \citep{Gieser2024} and the C$^{18}$O streamer in IRS3C (Fig. \ref{fig:streamers}, this work) are in agreement with the large scale rotation of the L1448N region (C$^{18}$O 30m clusters 1 and 2 and C$^{18}$O merged cluster 1 in Fig. \ref{fig:clusters}) with blueshifted and redshifted velocities towards the west and east, respectively. This implies that these streamers are infalling structures originating from the natal large-scale envelope material. In the streamline model it is explicitly assumed that the particle trajectory originates from the larger scale rotating envelope (Sect. \ref{sec:streamline}). By using the combination of large-scale single-dish maps and sensitive NOEMA observations, we confirm for the first time that this can be the case for observed streamers. Comparing our results with high resolution ALMA observations resolving the disk rotation \citep{Tobin2018,Reynolds2021}, the rotational motion is coherent from the large-scale envelope, to the infalling streamers, down to the disk scales. This implies that the protostellar systems have not formed in isolation, but originate from the same envelope material.
	
	The SO$_{2}$ streamer falling onto IRS3C (that is also bright in SO emission, Fig. \ref{fig:channel_map}) however does not follow the large-scale rotational pattern (Fig. \ref{fig:streamers}). The emission of SO and SO$_{2}$ is known to trace both accretion and ejection processes \citep{ArturdelaVillarmois2023} and SO is also prominent in the inner hot envelope \citep{Tychoniec2021}. Since the SO$_{2}$ line width is not broadened (Fig. \ref{fig:clusters}), the emission is likely not caused by a strong shock. It is more likely that the SO$_{2}$ streamer is impacted by the hotter inner region of IRS3C and additional interactions with the outflow material could be causing a change in the chemical composition with more abundant sulfur-bearing species. This streamer might originate from a cloudlet capture event, as it has been proposed for the streamer in DG Tau \citep{Hanawa2024}. A case of three infalling streamers has been found towards the source IRAS\,04239+2436 \citep{Lee2023} where the streamer spiral structures have been detected in SO emission, and one in addition in SO$_{2}$. Comparing the observational results with magneto-hydrodynamic simulations, the authors found that the streamers are consistent with gravitational interactions between the turbulent infalling envelope material and the central triple system and that the emission of S-bearing species is originating from low-velocity shocks from these interactions. Given that IRS3C is a binary system, this could also explain the SO$_{2}$ streamer. The streamers traced by C$^{18}$O and DCN in the region are on the other hand not bright in SO and SO$_{2}$. This could imply lower density and temperatures along these streamers and hence shocks might not play an important role. In contrast, the infall rate of the SO/SO$_{2}$ streamer is higher. This could cause shocks due to interaction with the envelope material and potentially even outflows.
	
	The large scale IRAM\,30m map of SO shows a different morphology compared to C$^{18}$O (Fig. \ref{fig:C18OSOSO2mom0}). Such differences between molecular emission following the dust continuum peak and molecular emission of SO and SO$_{2}$ showing asymmetries with respect to the dust continuum can be evident already in the prestellar stage, as it has been observed towards the prestellar core L1544 \citep{Spezzano2017}. The SO$_{2}$ streamer might be originating from that gas reservoir. To investigate this potential connection, deep observations of sulfur-bearing species with a larger FOV would be required.
	
\section{Conclusions}\label{sec:conclusions}

	In this study, we analyzed the kinematic properties of the L1448N region located in the Perseus molecular cloud combining single-dish observations with the IRAM\,30m telescope (ANTIHEROES) and interferometric observations with NOEMA (PRODIGE) tracing scales from 36\,000\,au down to 300\,au.
	
	\begin{enumerate}
		\item The C$^{18}$O emission traces the large-scale envelope material in the single-dish maps where we spectrally resolve large-scale rotation in which the three protostellar systems (IRS3A, IRS3B, and IRS3C) are embedded in. The higher resolution combined data resolves highly sub-structured material.
		\item Comparing NOEMA-only and merged (NOEMA+30m) C$^{18}$O data, we find that more than 90\% of the flux is filtered out in the interferometric data. With the exception of one SO$_{2}$ transition, all other lines analyzed in this work were heavily impacted by spatial filtering with more than 50\% missing flux. Not only the total flux, but also the line profiles are severely affected. Hence, the ANTIHEROES data provides invaluable information of all molecular lines with extended emission of the entire PRODIGE sample.
		\item Temperature maps using c-C$_{3}$H$_{2}$ lines indicate low temperatures ($<$20\,K) on average in the region in both the single-dish and merged data. Elevated temperatures are found in the merged data towards the location of the embedded protostars as well as towards the cavity walls of IRS3C.
		\item Towards IRS3C, we find two infalling streamers detected each in C$^{18}$O and sulfur-bearing species (SO, SO$_{2}$). The velocity profiles and trajectories of the infalling streamers found towards IRS3C are consistent with gravitational infall.
		\item The large-scale rotation of the L1448N envelope material traced by C$^{18}$O ($2-1$) is coherent down to the streamer kinematics as well as the disk rotation of the three protostellar systems. The infalling streamers in C$^{18}$O in IRS3C (this work) and DCN in IRS3A \citep{Gieser2024} accelerate towards the protostellar systems. These infalling streamers are consistent with remnant envelope material carved out by the bipolar outflows. The SO$_{2}$ streamer of IRS3C shows a different velocity pattern and might originate from an additional nearby gas reservoir that is detected in SO in the single-dish maps.
		\item The infall rates of the streamers falling onto IRS3A and IRS3C are $1\times10^{-6}$\,$M_\odot$\,yr$^{-1}$ and $5-18\times10^{-6}$\,$M_\odot$\,yr$^{-1}$, respectively. Thus, IRS3C, which is expected to be younger based on bolometric temperature, is accreting at higher accretion rates compared to IRS3A.
	\end{enumerate}
	
	In this work we find that streamers in young Class 0/I systems are connected to the remnant infalling envelope, while in the same system, asymmetric gas inflows with an origin that is not consistent with the overall envelope motion can occur at the same time. The ongoing ANTIHEROES program enables for the first time a concise characterization of the origin and chemical composition of infalling material in order to better constrain the nature and impact of streamers in a sample of Class 0/I systems in the Perseus star-forming region. Further sensitive high angular resolution observations of molecular line tracers are then necessary to spatially resolve the connection and impact of the infalling streamers to the disk properties.

\begin{acknowledgements}
	The authors thank the anonymous referee for their constructive report that helped improve the clarity of this paper. The authors thank the IRAM staff at the NOEMA observatory and IRAM 30m telescope for their support in the observations and data calibration. This work is based on observations carried out under project numbers L19MB and 096-23 with the IRAM NOEMA Interferometer and IRAM 30m telescope, respectively. IRAM is supported by INSU/CNRS (France), MPG (Germany) and IGN (Spain). This project is co-funded by the European Union (ERC, SUL4LIFE, grant agreement No101096293). A.F. also thanks project PID2022-137980NB-I00 funded by the Spanish
Ministry of Science and Innovation/State Agency of Research MCIN/AEI/10.13039/501100011033 and by “ERDF A way of making Europe”. D.~S. has received funding from the European Research Council (ERC) under the European Union’s Horizon 2020 research and innovation programme
(PROTOPLANETS, grant agreement No. 101002188).
\end{acknowledgements}

\bibliographystyle{aa}
\bibliography{bibliography}

\begin{appendix}

\onecolumn
\section{Observation properties}

	Table \ref{tab:obs} summarizes the transition properties and observation parameters of all molecular lines analyzed in this work. The data calibration and imaging procedures are described in detail in Sect. \ref{sec:observations}.

\setlength{\tabcolsep}{2pt}
\begin{table*}[!htb]
\caption{Molecular line properties and observational parameters of the PRODIGE-ANTIHEROES data.}
\label{tab:obs}
\centering
\renewcommand{\arraystretch}{1.2}
\begin{tabular}{lrrr | r | rr | rr | rr | r}
\hline\hline
 & & & & & \multicolumn{2}{c|}{\underline{IRAM\,30m}} & & & & & \\ 
Transition & Rest- & Upper & Einstein $A$ & Channel & Beam & Line & Beam & Line & Lower & Upper & Missing\\ 
 & frequency & energy & coefficient & width & & noise & & noise & \multicolumn{2}{c|}{velocity} & flux\\ 
 & $\nu$ & $E_\mathrm{u}/k_\mathrm{B}$ & log $A_\mathrm{ul}$ & $\delta \varv$ & $\theta$ & $\sigma_\mathrm{line,30m}$ & $\theta_\mathrm{maj}\times\theta_\mathrm{min}$ (PA) & $\sigma_\mathrm{line}$ & $\varv_\mathrm{low}$ & $\varv_\mathrm{upp}$ & \\
 & (GHz) & (K) & (log s$^{-1}$) & (km\,s$^{-1}$) & ($''$) & (mK) & ($''\times''$($^\circ$)) & (K) & \multicolumn{2}{c|}{(km\,s$^{-1}$)} & (\%) \\
\hline
 & & & & & & & \multicolumn{2}{c|}{\underline{NOEMA+IRAM\,30m}} & & & \\ 
c-C$_{3}$H$_{2}$\,($3_{3,0}-2_{2,1}$) & 216.279 & 19.5 & $-3.6$ & 0.1 & 12.0 & 99 & 1.28$\times$0.92 (15) & 0.34 & $3.0$ & $7.0$ & 98\\ 
DCN\,($3-2$) & 217.239 & 20.9 & $-3.3$ & 0.1 & 11.9 & 100 & 1.28$\times$0.92 (15) & 0.35 & $1.0$ & $10.0$ & 57\\ 
c-C$_{3}$H$_{2}$\,($6_{0,6}-5_{1,5}$) & 217.822 & 38.6 & $-3.3$ & 0.1 & 11.9 & 93 & 1.27$\times$0.92 (15) & 0.35 & $3.0$ & $7.0$ & 85 \\ 
c-C$_{3}$H$_{2}$\,($5_{1,4}-4_{2,3}$) & 217.940 & 35.4 & $-3.4$ & 0.1 & 11.9 & 96 & 1.27$\times$0.91 (15) & 0.35 & $3.0$ & $7.0$ & 88\\ 
C$^{18}$O\,($2-1$) & 219.560 & 15.8 & $-6.2$ & 0.1 & 11.8 & 101 & 1.26$\times$0.91 (14) & 0.34 & $1.0$ & $11.0$ & 93\\ 
SO\,($6_{5}-5_{4}$) & 219.949 & 35.0 & $-3.9$ & 0.1 & 11.8 & 101 & 1.26$\times$0.91 (14) & 0.38 & $2.0$ & $6.0$ & 73\\ 
CO\,($2-1$) & 230.538 & 16.6 & $-6.2$ & 0.1 & 11.2 & 99 & 1.22$\times$0.85 (13) & 0.38 & $\ldots$ & $\ldots$ & 90\\ \hline
 & & & & & & & \multicolumn{2}{c|}{\underline{NOEMA}} & & \\ 
SO$_{2}$\,($4_{2,2}-3_{1,3}$) & 235.152 & 19.0 & $-4.1$ & 0.1 & 11.0 & 106 & 1.16$\times$0.81 (15) & 0.39 & $2.0$ & $6.0$ & \ldots \\ 
\hline
\end{tabular}
\tablefoot{The transition properties (quantum numbers, $\nu$, $E_\mathrm{u}$/$k_\mathrm{B}$, and $A_\mathrm{ul}$) are taken from the Cologne Database for Molecular Spectroscopy \citep[CDMS,][]{CDMS}. The noise per channel in the line data, $\sigma_\mathrm{line}$, is estimated in the central area of the mosaic as the standard deviation in line-free channels. The last two columns show the velocity ranges, $\varv_\mathrm{low}$ and $\varv_\mathrm{upp}$, used to compute the line integrated intensity maps. The estimate of missing flux due to spatial filtering is calculated based on an integrated intensity comparison considering all areas with S/N $>$5 in the NOEMA and merged data.}
\end{table*}

\section{Comparison of DCN ($3-2$) and C$^{18}$O ($2-1$) NOEMA-only and merged data products}\label{sec:app_merging}

\begin{figure*}[!htb]
\centering
\includegraphics[width=0.49\textwidth]{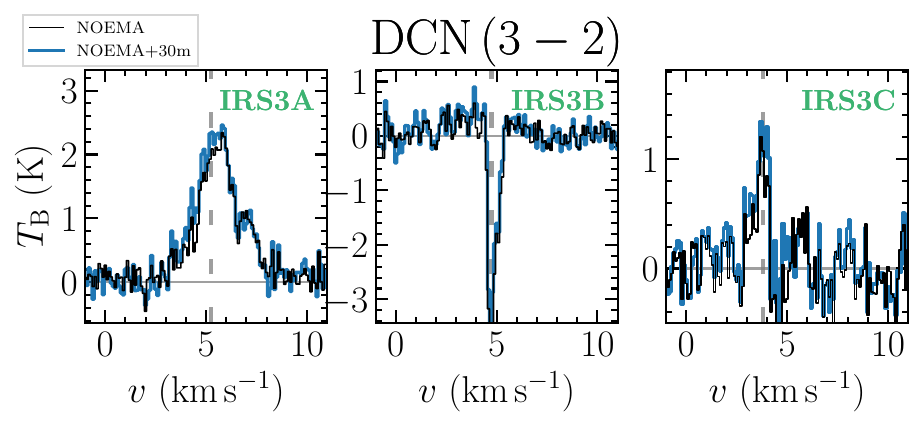}
\includegraphics[width=0.49\textwidth]{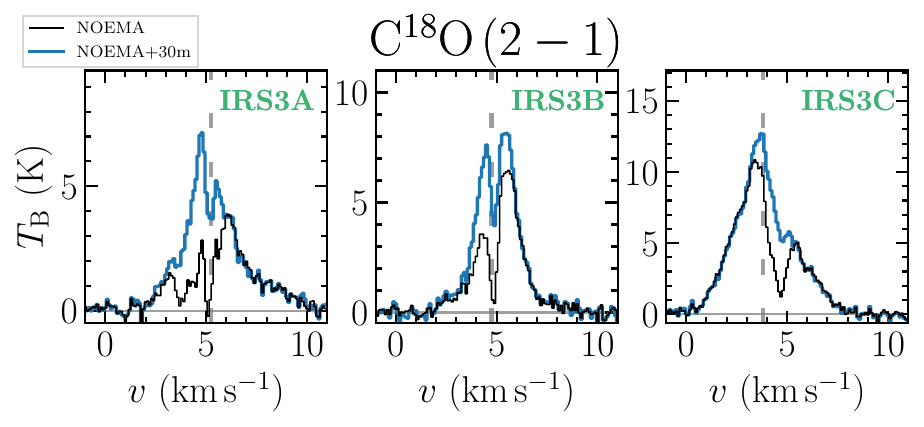}
\caption{Spectra of DCN $3-2$ (left) and C$^{18}$O $2-1$ (right) extracted from the continuum peak positions of IRS3A, IRS3B, and IRS3C. In black the NOEMA only spectrum is shown, while in blue the merged (NOEMA+30m) data are presented. The source $\varv_\mathrm{LSR}$ is indicated by the vertical dashed line.}
\label{fig:comparison_spectra}
\end{figure*}

\begin{figure*}[!htb]
\centering
\includegraphics[width=0.9\textwidth]{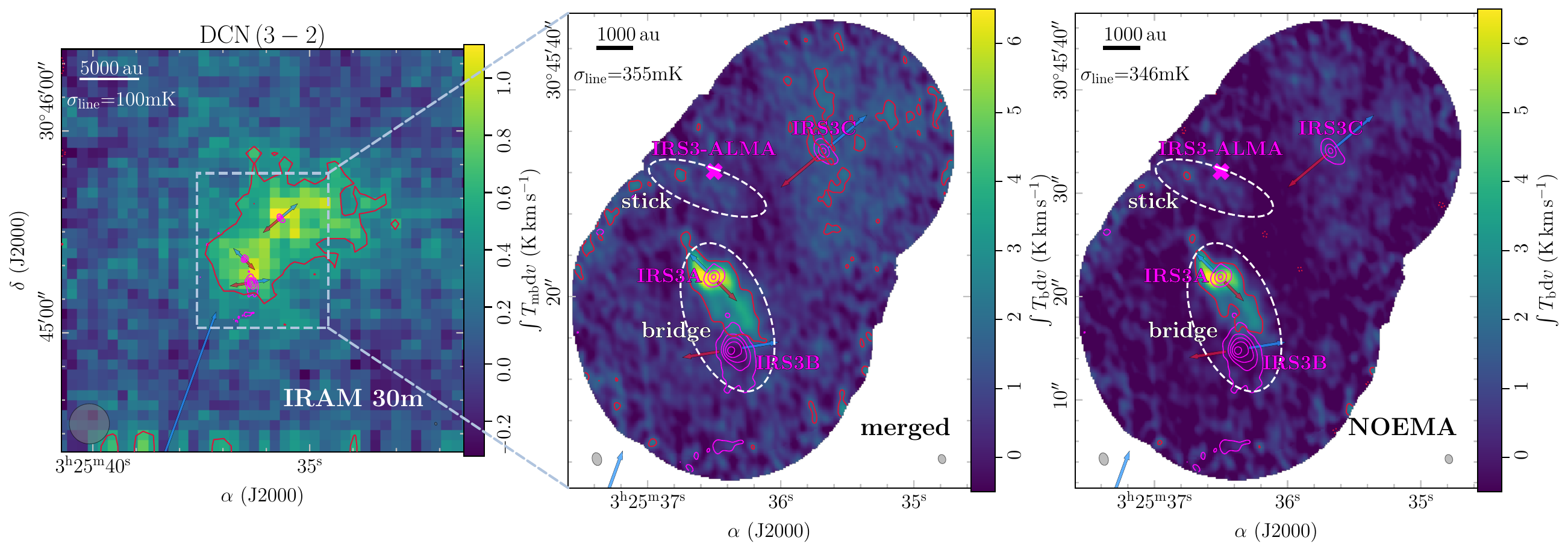}
\includegraphics[width=0.9\textwidth]{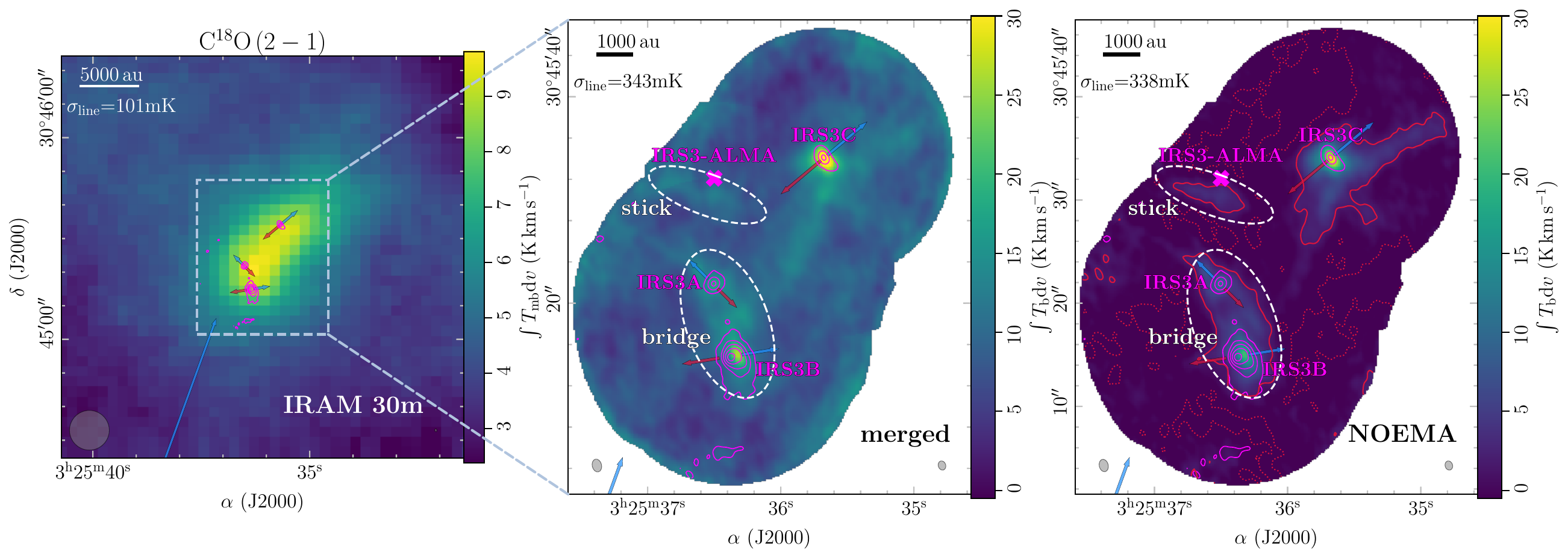}
\caption{Line integrated intensity map of DCN ($3-2$) and C$^{18}$O ($2-1$) of the IRAM\,30m data (left) and merged data (center), and NOEMA data (right). Red contours highlight emission of $-$5$\sigma$ (dotted) and 5$\sigma$ (solid). If no contours are shown, the emission is $>5\sigma$ in the entire FOV. Pink contours and arrows are the same as in Fig. \ref{fig:continuum}. The synthesized beam of the line and continuum data is shown in the bottom left and right corner, respectively. Scale bars are shown in the top left corner.}
\label{fig:comparison}
\end{figure*}

	Figure \ref{fig:comparison_spectra} shows a comparison between DCN ($3-2$) and C$^{18}$O ($2-1$) spectra extracted from the continuum peak positions of IRS3A, IRS3B and IRS3C using the NOEMA-only data as well as the merged data. Line integrated intensity maps of both transitions comparing the single-dish, merged, and NOEMA-only data are shown in Fig. \ref{fig:comparison}.
	
	The spatial distribution of DCN is mostly compact tracing the gas bridge towards IRS3A and IRS3B \citep{Gieser2024}, but the additional IRAM 30m short spacing information reveals weak extended emission around IRS3C. This extended emission surrounding IRS3C accounts for in total 57\% missing flux in DCN ($3-2$) considering all areas with emission $>5\sigma$ in the line integrated intensity maps (Fig. \ref{fig:comparison}). The kinematic properties of the gas bridge analyzed in \citet{Gieser2024} using the DCN ($3-2$) transition is not affected by spatial filtering with similar intensities measured in the merged and NOEMA data towards IRS3A and IRS3B.
	
	The emission of C$^{18}$O ($2-1$) however is very extended being detected in the entire single-dish map. Comparing the merged and NOEMA-only integrated intensity maps stresses that while compact structures are recovered by NOEMA, most of the extended emission is not. In the C$^{18}$O ($2-1$) NOEMA data the missing flux is in total 93\%. While the integrated intensity maps of the NOEMA data have large areas with negative values, adding additional 30m data removes these artifacts completely. 
	
	In Table \ref{tab:obs} the percentage of missing flux is summarized for all transitions with extended emission used in this work. The amount of missing flux depends heavily on the transition and varies between 57\% and 98\%. Both the total flux Fig. \ref{fig:comparison}) as well as velocity profiles (Fig. \ref{fig:comparison_spectra}) of the C$^{18}$O ($2-1$) are different in the merged and NOEMA-only data. Hence, the NOEMA-only data should not be used in the case of molecular emission with extended emission when the analysis relies total flux measurements and/or kinematic properties. 

\section{Details of the kinematic analysis}\label{app:kinematic}

\begin{figure*}[!htb]
\centering
\includegraphics[width=0.9\textwidth]{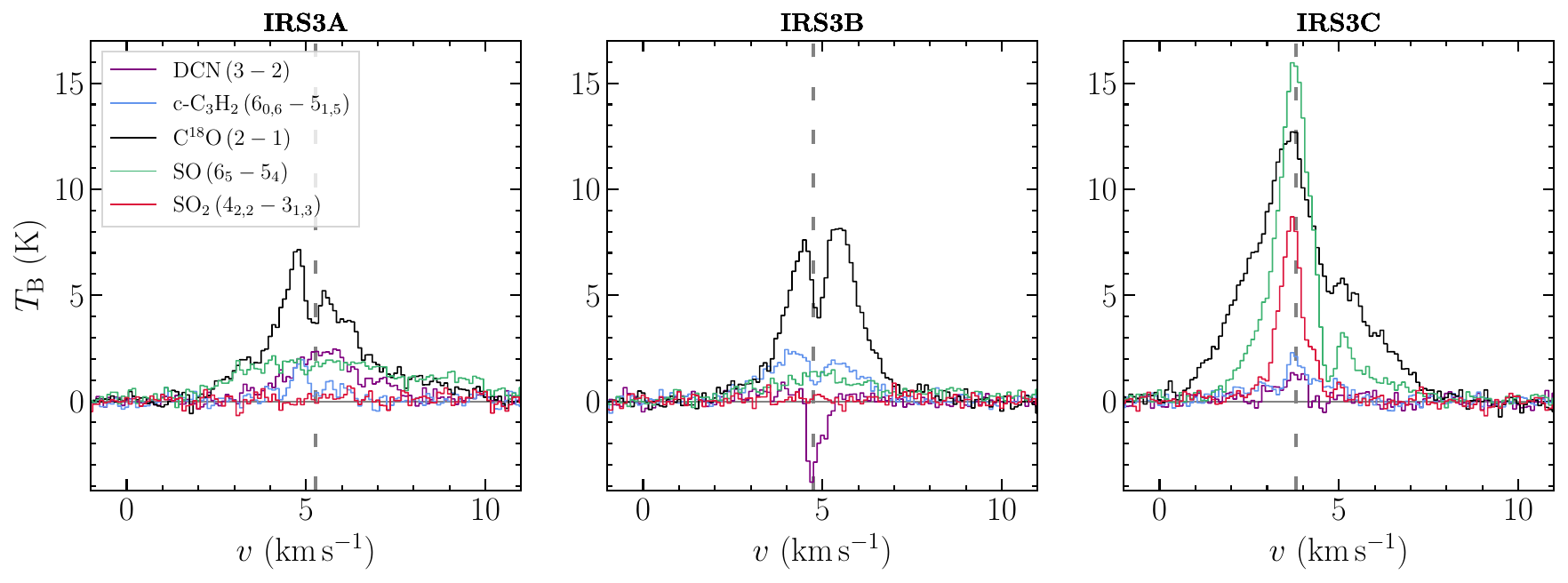}
\caption{Molecular line spectra extracted from the continuum peak positions of IRS3A (left), IRS3B (center), and IRS3C (right). The source $\varv_\mathrm{LSR}$ is indicated by the vertical dashed line. For SO$_{2}$ the NOEMA-only data, while for the remaining transitions the merged (NOEMA+IRAM 30m) data is shown (Table \ref{tab:obs}).}
\label{fig:spectra}
\end{figure*}

\setlength{\tabcolsep}{4pt}
\begin{table*}[!htb]
\caption{Overview of the kinematic analysis.}
\label{tab:kinematics}
\centering
\renewcommand{\arraystretch}{1.1}
\begin{tabular}{ll|rrr|rr|rrrr}
\hline\hline
 & & \multicolumn{3}{c}{\underline{Quality assessment}} & \multicolumn{2}{c}{\underline{Clustering setup}} & \multicolumn{4}{c}{\underline{Clustering results}}\\
Molecule & Data & Maximum & Maximum & AIC & \multicolumn{2}{c|}{\texttt{DBSCAN}} & Total & Input & Averaged & Discarded \\ 
 & product & noise & uncertainty & criterion & \multicolumn{2}{c|}{parameters} & clusters & \multicolumn{3}{c}{data points}\\ 
 & & $\sigma_{line,max}$ & ($I_\mathrm{peak}$, $\varv$, or $\Delta \varv$) & $\Delta$AIC & \texttt{eps} & \texttt{minsamples} & & & & \\ 
C$^{18}$O & 30m & 0.13\,K & 10\% & 30 & 0.28 & 50 & 3 & 1983 & 27 & 182 \\
C$^{18}$O & merged & 0.28\,K & 10\% & 30 & 0.24 & 230 & 9 & 26\,927 & 310 & 2641 \\
SO$_{2}$ & NOEMA & 0.4\,K & 30\% & 10 & 0.31 & 30 & 3 & 1347 & 140 & 161 \\
\hline
\end{tabular}
\tablefoot{Averaged data points are the number of positions where \texttt{DBSCAN} assigned multiple velocity components at the same position to a cluster. In these cases the parameters of the velocity components were averaged. All data points that were not assigned to any cluster by \texttt{DBSCAN} are referred to as discarded data points.}
\end{table*}

	Figure \ref{fig:spectra} shows spectra extracted from IRS3A, IRS3B, and IRS3C continuum peak positions comparing the line profiles of DCN ($3-2$), c-C$_{3}$H$_{2}$ ($6_{0,6}-5_{1,5}$), C$^{18}$O ($2-1$), SO ($6_5-5_4$), and SO$_{2}$ ($4_{2,2}-3_{1,3}$). The strong line absorption of DCN and C$^{18}$O towards IRS3B is due to absorption against the bright continuum background emission. The high spectral resolution of the molecular line data allows a detailed analysis of the kinematic properties of the gas conducted in Sect. \ref{sec:kinematics}. Here, a detailed description of the methodology based on the analysis of IRS3A \citep{Gieser2024} is presented. 
	
	In Sect. \ref{sec:kinematics} we perform a Gaussian decomposition of C$^{18}$O (2-1) and SO$_{2}$ ($4_{2,2}-3_{1,3}$) where up to three velocity components are fitted in each spectrum. Given the increase of noise towards the edges of the FOV, the noise (per 0.1\,km\,s$^{-1}$ channel) is estimated in each spectrum individually. The minimum amplitude of each Gaussian velocity component is set to at least 5$\times$ the noise and the minimum full-width at half maximum (FWHM) line width is set to 0.1\,km\,s$^{-1}$. We discard all pixels with a defined maximum noise value $\sigma_\mathrm{line,max}$ to mask noisy areas at the edges of the FOV.
	
	We evaluate the goodness of the one-, two-, and three-component fits in each spectrum based on the Akaike Information Criterion (AIC). The best-fit is selected based on the lowest AIC value, but only if the difference to the second lowest value is higher than $\Delta$AIC, defined for each transition. We perform a second quality assessment where we discard all velocity components which have relative uncertainties in any of the fit parameters (peak velocity $\varv$, FWHM line width $\Delta \varv$, amplitude $I_\mathrm{peak}$) higher than a defined percentage. The quality assessment of the fitted Gaussian velocity components is summarized in Table \ref{tab:kinematics} for all analyzed data sets. 
	
	Given the extended emission and complex spectral line profiles of C$^{18}$O in both the single-dish and merged data, our criteria are more conservative compared to SO$_{2}$, in order to only extract the most reliable velocity components. We note that in the C$^{18}$O merged data due to our conservative approach many velocity components are discarded after the quality assessment, even though line emission may be bright, but in many cases line profiles are too complex to fit them properly. This is especially the case towards the complex bridge region surrounding IRS3A and IRS3B where in addition to the infalling gas bridge, there is a significant contribution from outflows and rotation of the compact envelopes and disks (Fig. \ref{fig:spectra}).

\begin{figure*}[!htb]
\centering
\includegraphics[width=0.44\textwidth]{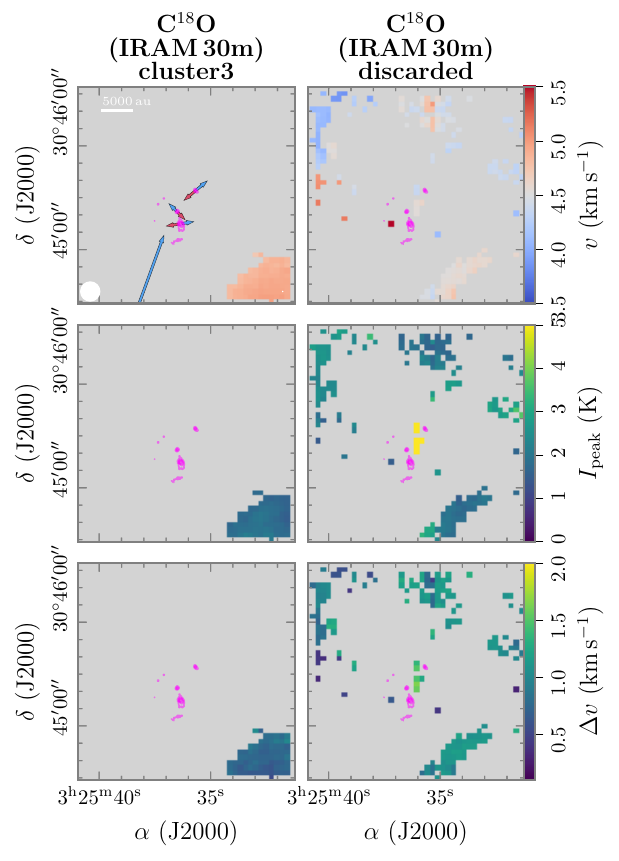}
\includegraphics[width=0.45\textwidth]{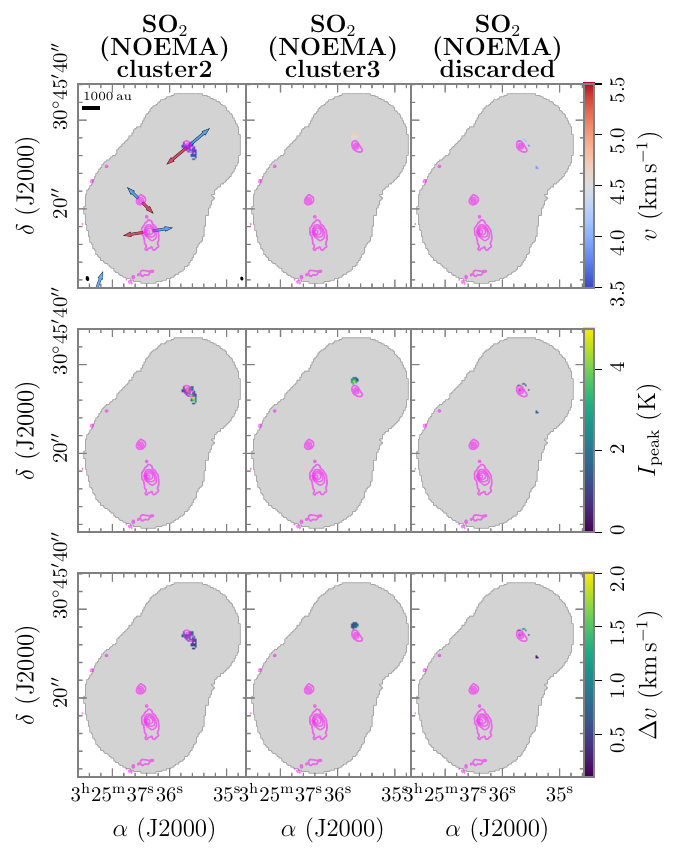}
\includegraphics[width=0.95\textwidth]{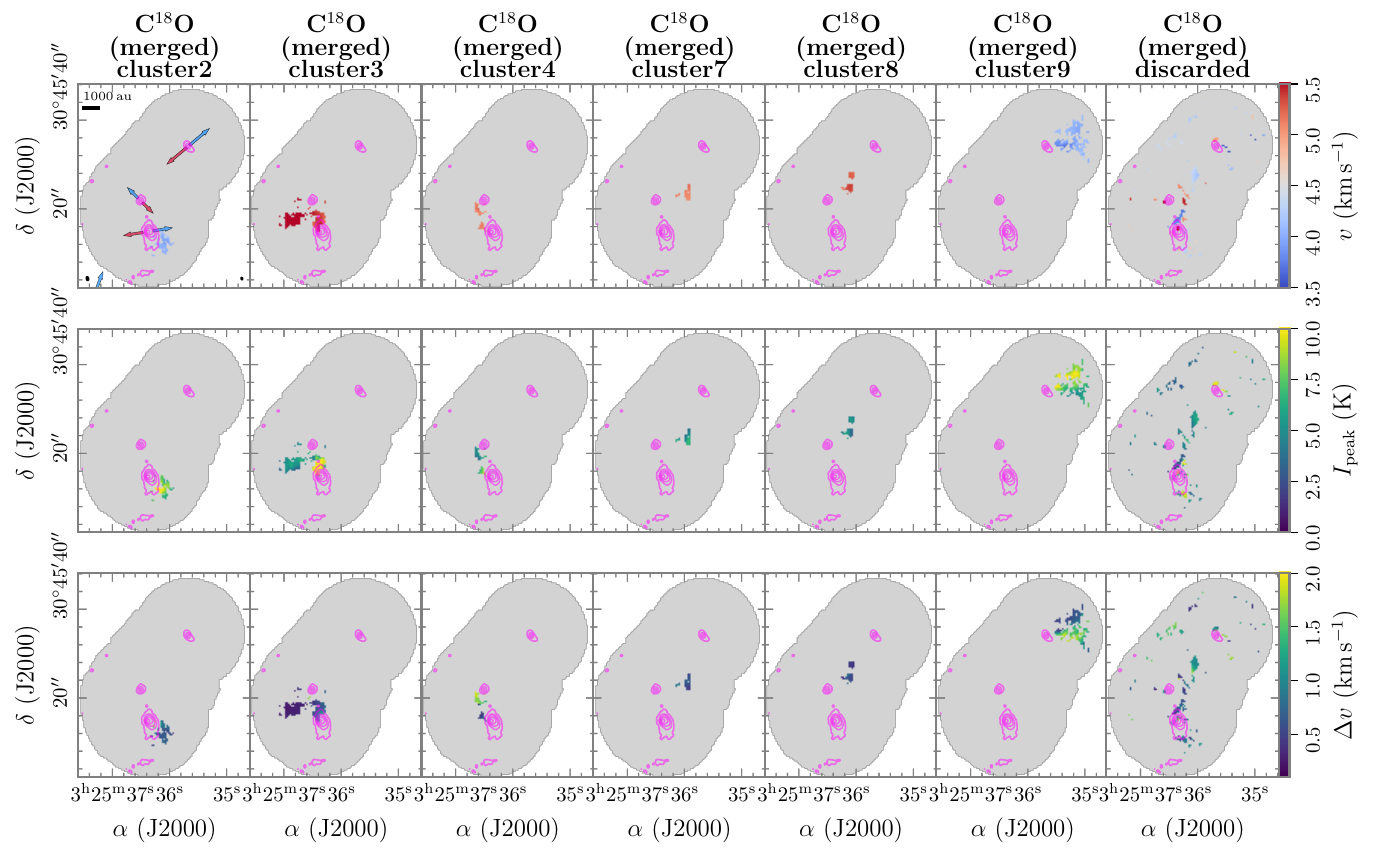}
\caption{The same as Fig. \ref{fig:clusters} presenting the main clusters, while here remaining clusters extracted with \texttt{DBSCAN} are shown.}
\label{fig:clusters_app}
\end{figure*}

	The data set with up to three velocity components per spectrum is sorted using the clustering algorithm \texttt{DBSCAN}. The clustering in \texttt{DBSCAN} is set by two parameters, the neighborhood parameter \texttt{eps} ($\varepsilon$) and the minimum number of data points within a cluster \texttt{minsamples}. For all data sets we check the parameter space of these two parameters and find a combination of the two that 1) creates smooth velocity maps in each cluster that also produce coherent amplitude and line width maps, 2) has a low number of discarded points, 3) is able to disentangle the velocity components at a certain position, so averaging of the data points within a single pixel is minimized. The input parameters, as well as clustering results are summarized in Table \ref{tab:kinematics}. Additional clusters as well as the discarded data points extracted with \texttt{DBSCAN} in Sect. \ref{sec:kinematics} are summarized in Fig. \ref{fig:clusters_app}.
	
	For example, large-scale redshifted emission in C$^{18}$O single-dish data (C$^{18}$O (IRAM\,30m) cluster 3) is most likely connected to a background cloud component, as towards this direction there is no known outflow. The disk rotation of IRS3C is extracted in SO$_{2}$ (NOEMA) cluster 2. The red-shifted SO$_{2}$ blob (SO$_{2}$ (NOEMA) cluster 3) towards the northwest might be connected to the arc structure (SO$_{2}$ (NOEMA) cluster 1), i.e., the arc is wrapping around behind the source where it is infalling. In the merged C$^{18}$O data set, many redshifted blobs are extracted in four clusters, oriented along the redshifted outflow lobe launched by IRS3C (C$^{18}$O (merged) clusters 3, 4, 7, and 8). Blueshifted outflow components are seen as well for IRS3B (C$^{18}$O (merged) cluster 2) and IRS3C (C$^{18}$O (merged) cluster 9).

\section{Excitation conditions of c-C$_{3}$H$_{2}$}\label{sec:app_linemodeling}

\begin{figure*}[!htb]
\centering
\includegraphics[width=0.8\textwidth]{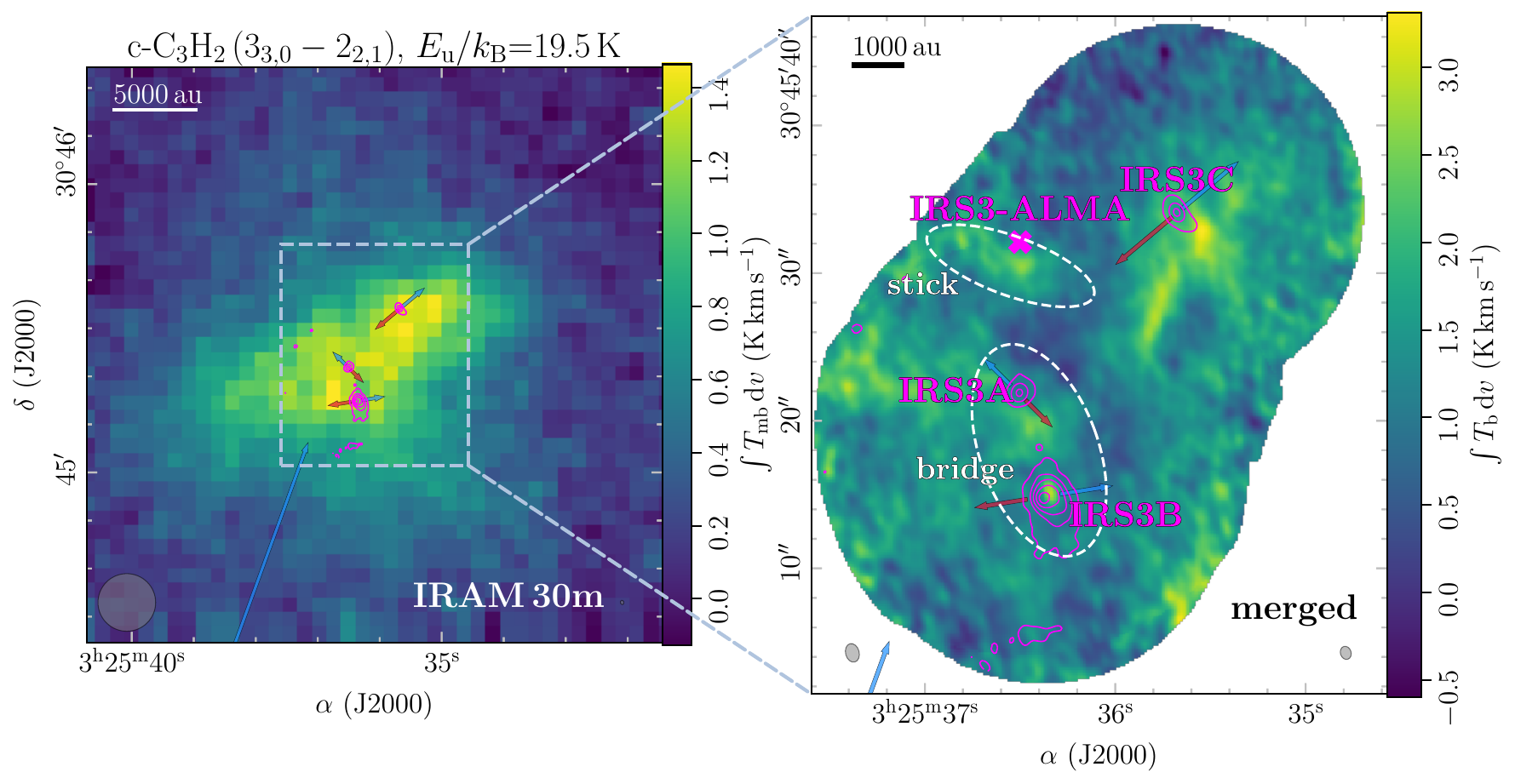}
\includegraphics[width=0.8\textwidth]{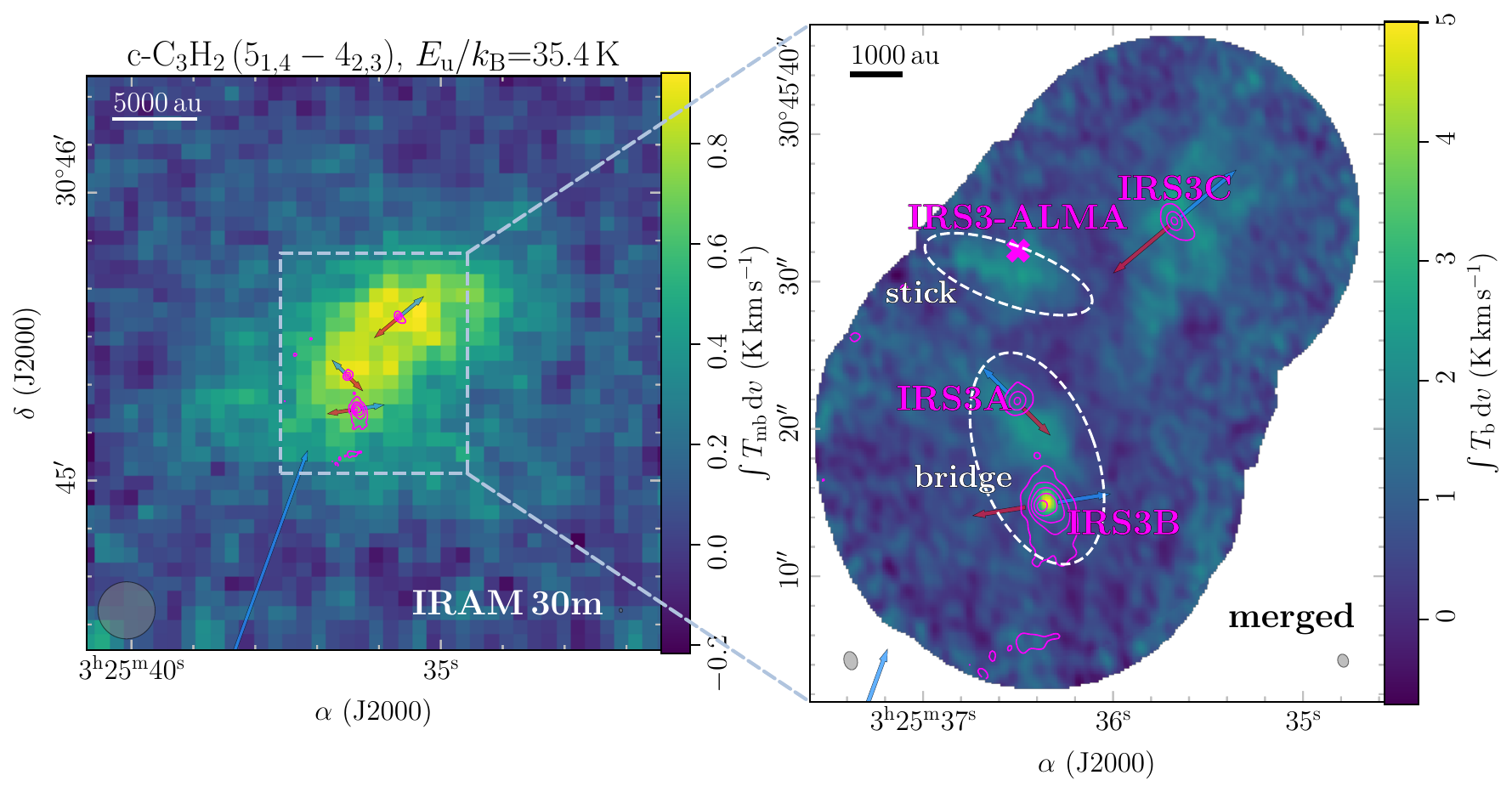}
\includegraphics[width=0.8\textwidth]{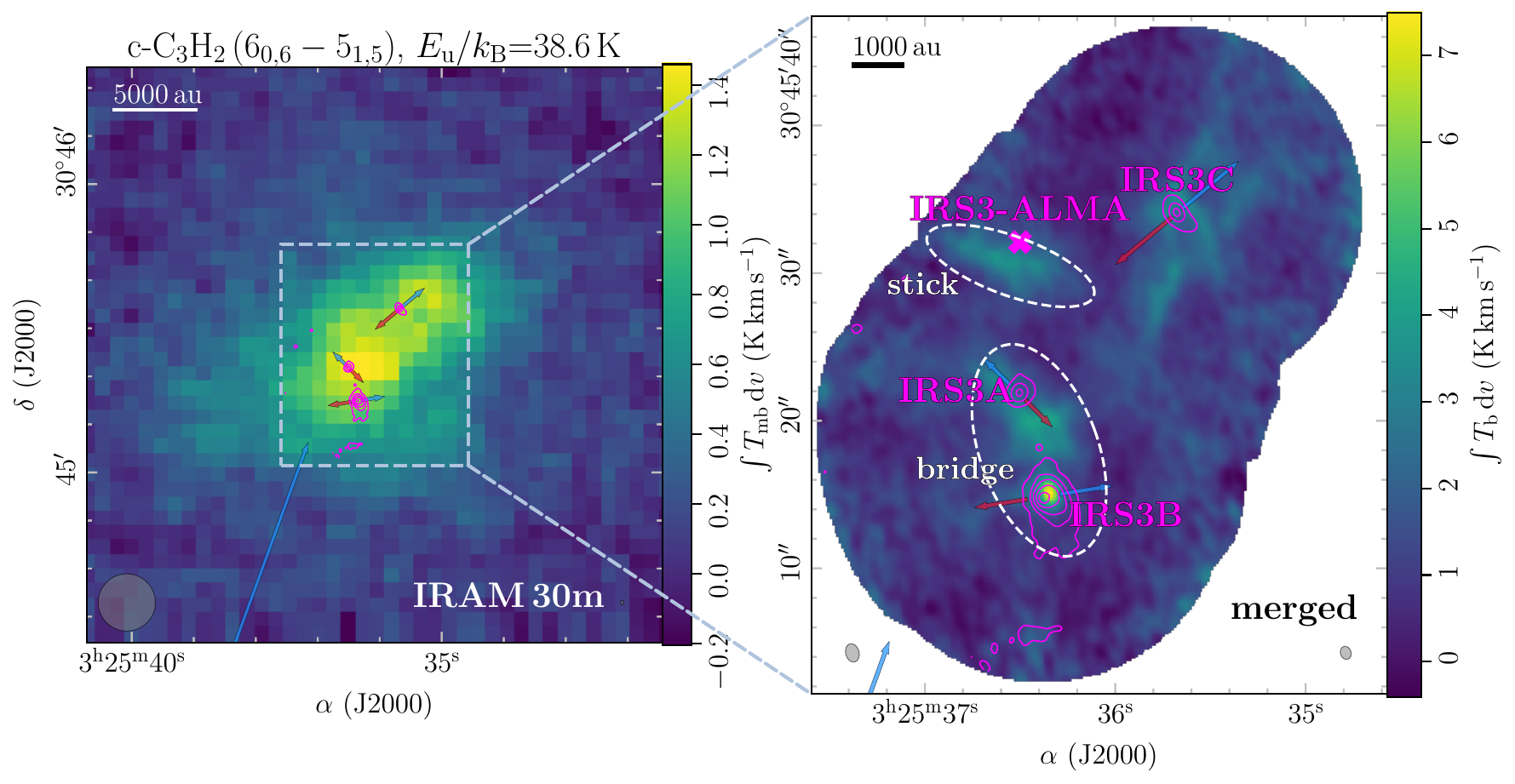}
\caption{The same as Fig. \ref{fig:C18OSOSO2mom0}, but for c-C$_{3}$H$_{2}$ transitions.}
\label{fig:mom0_C3H2}
\end{figure*}

	Integrated intensity maps of all three detected c-C$_{3}$H$_{2}$ transitions are shown in Fig. \ref{fig:mom0_C3H2}. In Sect. \ref{sec:radtrans} we use the \texttt{XCLASS} tool to derive the rotation temperature and column density maps. The detailed setup of the line fitting with \texttt{XCLASS} is described in the following.
	
		Since the noise is not homogeneous and in general increases towards the edges of the FOV, we evaluate the noise in each pixel individually. The noise is computed from the average noise calculated in line-free velocity ranges on the blue- and red-shifted sides of each emission line. All pixels with noise values higher than 0.15\,K and 0.4\,K in the 30m and merged data, respectively, are flagged which correspond to areas at the edge of the FOV. We then compute S/N maps of all transitions and only spectra where the S/N is larger than 6 for all three transitions are fitted with \texttt{XCLASS}.
	
	The \texttt{myXCLASSFit} function solves the radiative transfer equation of an isothermal source for all selected transitions of a molecule assuming local thermal equilibrium (LTE) conditions. The critical densities for the three c-C$_{3}$H$_{2}$ transitions range between $4-5\times10^7$\,cm$^{-2}$ \citep[using collisional rates at 30\,K from the Leiden Atomic and Molecular Database,][]{vanderTak2020}. In a spherical symmetric envelope approximation, at distances of several hundreds of au from the protostar, densities can reach such high densities \citep{Gerin2017}. At larger distances, the underlying kinetic temperature might be slightly underestimated from this approach. The fit parameter set includes the source size $\theta_\mathrm{source}$, rotation temperature $T_\mathrm{rot}$, column density $N$, FWHM line width $\Delta \varv$, and peak velocity $\varv$.
	
	To reduce the fit parameter set, we fix the source size $\theta_\mathrm{source}$ to 12$''$ and 2$''$ for all 30\,m and merged data, respectively. This implies a beam filling factor of 1, which is reasonable since the emission is extended for all c-C$_{3}$H$_{2}$ transitions (Fig. \ref{fig:mom0_C3H2}). The best-fit model (based on the lowest reduced $\chi^2$) is determined by using an algorithm chain starting with the Genetic algorithm (100 iterations) followed by the Levenberg-Marquart algorithm (50 iterations) optimizing the parameter set towards global and local minima, respectively.

\section{Column density maps of the envelope and the streamers}\label{app:streamer_coldens}

\begin{figure*}[!b]
\centering
\includegraphics[width=0.4\textwidth]{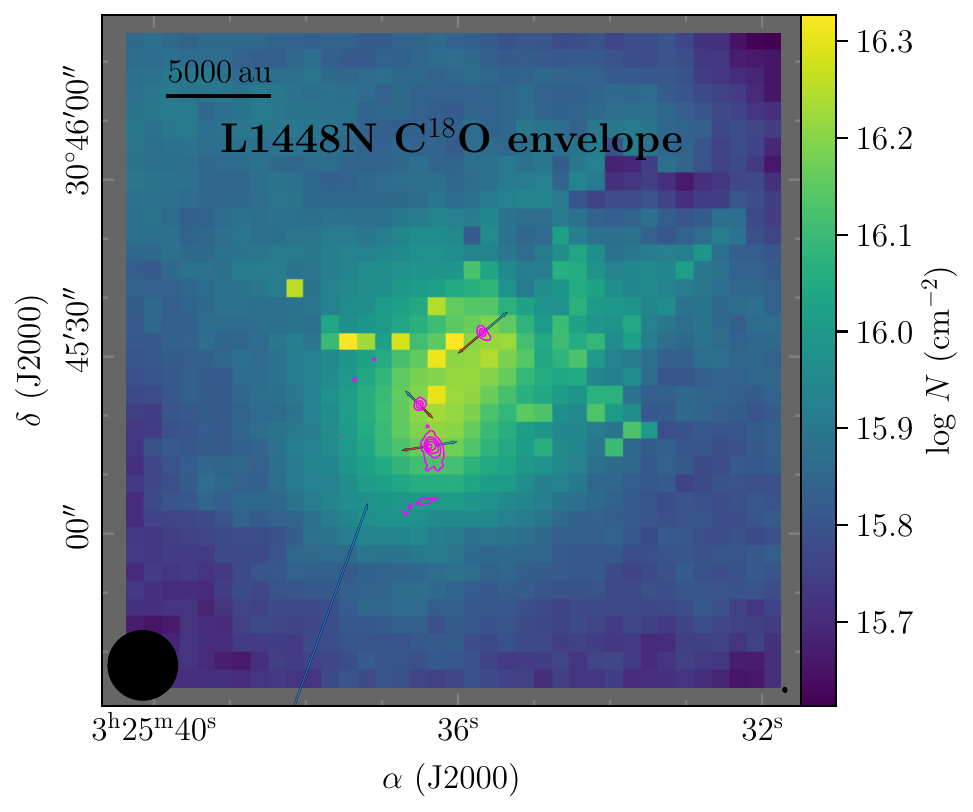}\\
\includegraphics[width=0.4\textwidth]{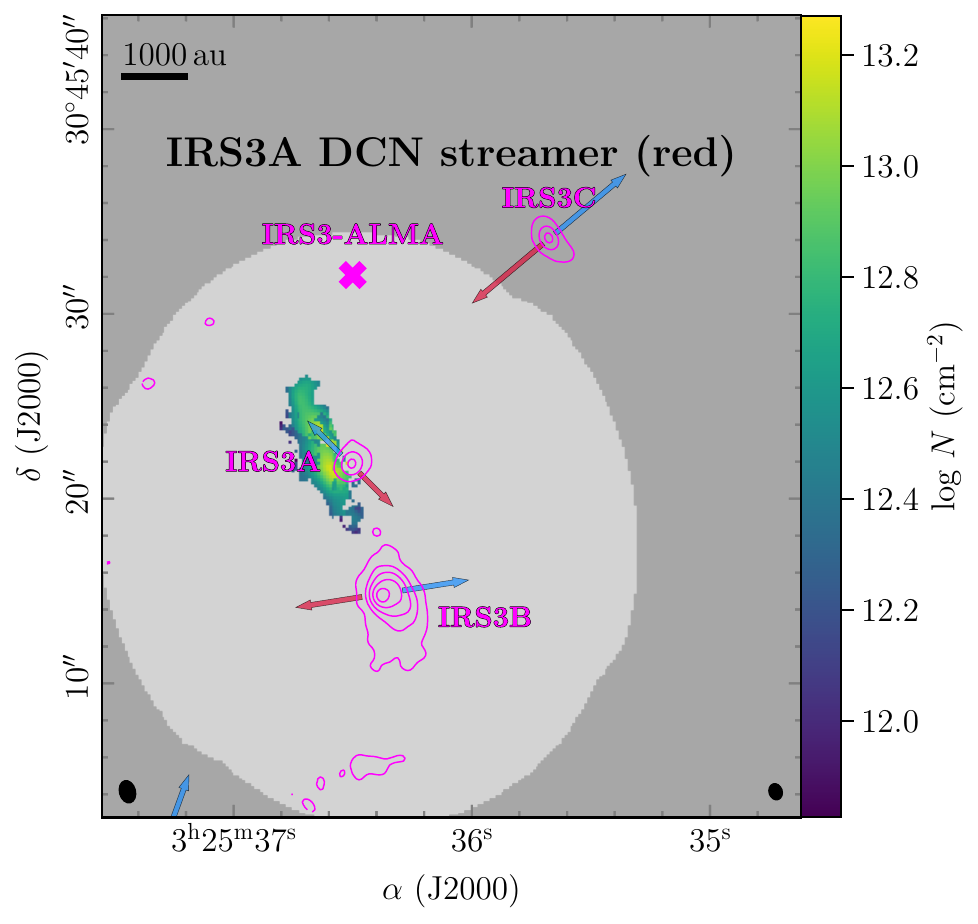}
\includegraphics[width=0.4\textwidth]{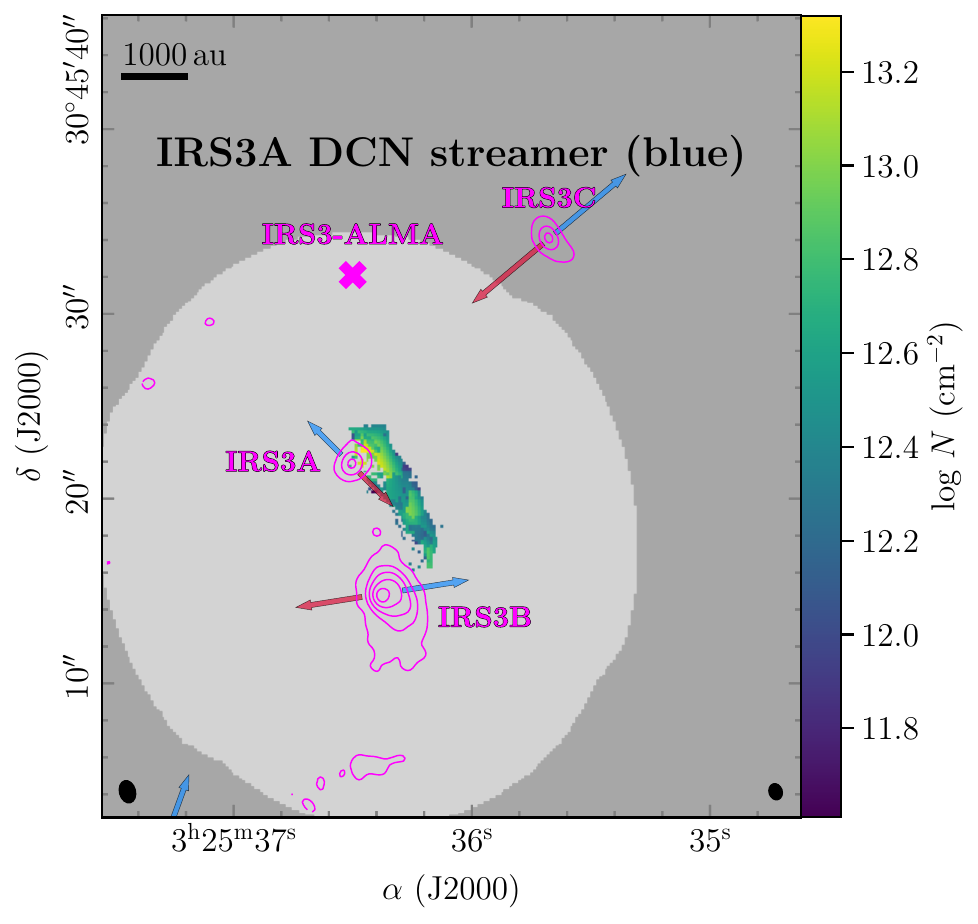}
\includegraphics[width=0.4\textwidth]{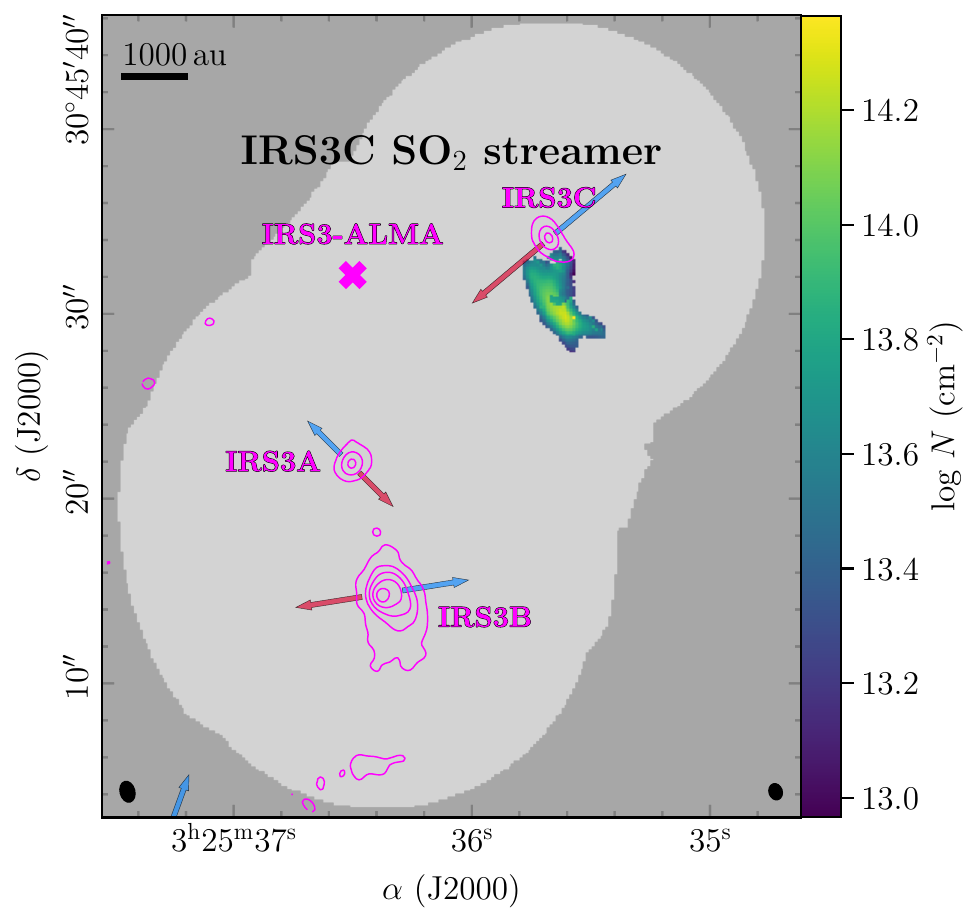}
\includegraphics[width=0.4\textwidth]{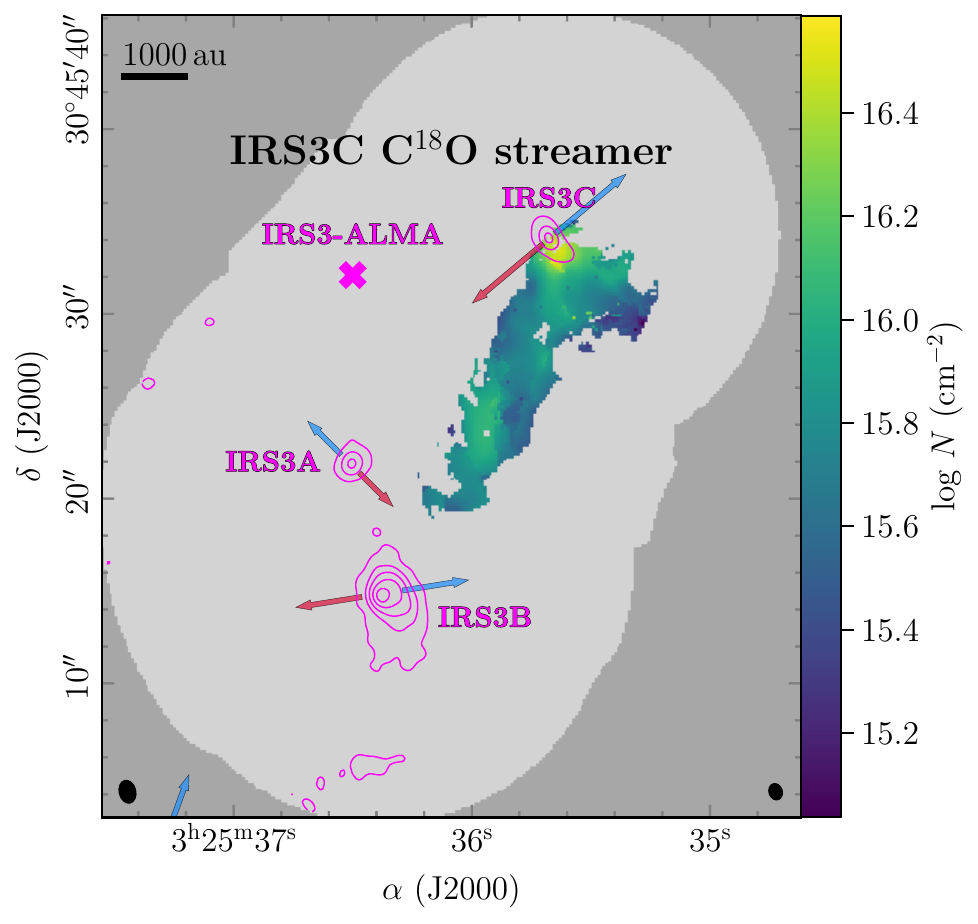}
\caption{Column density maps of the L1448N and the streamers. Contours and arrows are the same as in Fig. \ref{fig:continuum}. The synthesized beam of the line and continuum data is shown in the bottom left and right, respectively. A scale bar is indicated in the top left corner. The IRS3A DCN streamers were analyzed in \citet{Gieser2024} and IRS3C C$^{18}$O and SO$_{2}$ streamers were analyzed in this work.}
\label{fig:coldens}
\end{figure*}

	Column density maps of the L1448N envelope as well as the streamers infalling onto IRS3A and IRS3C are presented in Fig. \ref{fig:coldens} and are used to estimate the masses in Sect. \ref{sec:infallrates}.

\end{appendix}
\end{document}